\documentclass[conference,compsoc]{IEEEtran}
\ifCLASSOPTIONcompsoc
  \usepackage[nocompress]{cite}
\else
  \usepackage{cite}
\fi
\ifCLASSINFOpdf
\else
\fi
\usepackage{amssymb}
\usepackage{lipsum}
\usepackage{tikz}
\usepackage[table]{xcolor}
\definecolor{lightgray}{gray}{0.9}
\usepackage{amsmath}
\usepackage{oplotsymbl}
\usepackage{filecontents}
\usepackage{pifont}
\usepackage[T1]{fontenc}
\usepackage{fontawesome5} 
\usepackage{comment}
\usepackage{wasysym}
\usepackage{multirow, hhline}
\usepackage{colortbl}

\usepackage{adjustbox}
\usepackage{graphicx} 
\usepackage{booktabs}
\usepackage{makecell}
\usepackage{utfsym}
\usepackage{twemojis}
\usepackage{tablefootnote}
\usepackage{subcaption}
\usepackage{url}
\usepackage{hyperref}
\usepackage{array} 
\usepackage{enumitem}
\usepackage[most]{tcolorbox}

\begin{document}
%
\title{SoK: The Next Frontier in AV Security: Systematizing Perception Attacks and the Emerging Threat of Multi-Sensor Fusion}


\author{\IEEEauthorblockN{Shahriar Rahman Khan}
\IEEEauthorblockA{Dept. of Computer Science\\
Kent State University\\
Kent, OH 44242, USA\\
srahmank@kent.edu}
\and
\IEEEauthorblockN{Tariqul Islam}
\IEEEauthorblockA{Dept. of Information Systems\\
University of Maryland, Baltimore County,\\
Baltimore, MD 21250, USA\\
mtislam@umbc.edu}
\and
\IEEEauthorblockN{Raiful Hasan}
\IEEEauthorblockA{Dept. of Computer Science\\
Kent State University\\
Kent, OH 44242, USA\\
rhasan7@kent.edu}}



%


\maketitle

\begin{abstract}
Autonomous vehicles (AVs) increasingly rely on multi-sensor perception pipelines that combine data from cameras, LiDAR, radar, and other modalities to interpret the environment. This SoK systematizes 48 peer-reviewed studies on perception-layer attacks against AVs, tracking the field’s evolution from single-sensor exploits to complex, cross-modal threats that compromise Multi-Sensor Fusion (MSF). We develop a unified taxonomy of 20 attack vectors organized by sensor type, attack stage, medium, and perception module, revealing patterns that expose underexplored vulnerabilities in fusion logic and cross-sensor dependencies. Our analysis identifies key research gaps including limited real-world testing, short-term evaluation bias, and the absence of defenses that account for inter-sensor consistency. To illustrate one such gap, we validate a fusion-level vulnerability through a proof-of-concept simulation combining infrared and LiDAR spoofing. The findings highlight a fundamental shift in AV security: as systems fuse more sensors for robustness, attackers exploit the very redundancy meant to ensure safety. We conclude with directions for fusion-aware defense design and a research agenda for trustworthy perception in autonomous systems.
\end{abstract}



%
\IEEEpeerreviewmaketitle

\section{Introduction}  
\label{sec:intro}  
AVs are transitioning from experimental prototypes to commercial and consumer realities, as evidenced by the increasing deployment of advanced driver-assistance systems (ADAS) and fully autonomous platforms in real-world environments~\cite{biswas2023autonomous, bagloee2016autonomous,fagnant2015preparing}. Major industry players including Tesla, Waymo, and Uber have accelerated commercial rollouts, with driverless taxi services now operating in multiple metropolitan areas. Additionally, autonomous buses, trucks, and delivery vehicles are now commercially available. However, as these vehicles become more widespread, incidents and accidents involving them have also increased. As of May 2024, a total of 617 AV crashes have been reported to NHTSA since July 2021, with 91 of those occurring in 2024 alone~\cite{nhtsa2024sgo}.

The AV system relies heavily on various sensors such as LiDAR, cameras, radar, and GPS to enable perception, localization, planning, and control \cite{li2025multi,zhou2023sensor}. Among these, perception serves as the first and most critical stage, where raw sensor data are interpreted to recognize and understand the driving environment. Consequently, any attack on sensor data can directly compromise perception output and cascade through the AV pipeline. While multi-sensor systems aim to mitigate the limitations of individual sensors, they also introduce new attack surfaces, allowing adversaries to manipulate sensor inputs and deceive perception models. For instance, spoofing LiDAR point clouds \cite{bhupathiraju2023emi}, injecting adversarial patches into camera feeds \cite{man2020ghostimage}, manipulating raw radar input using radio signals \cite{sun2021control, guesmi2022adversarial} or fabricating collaborative sensor data \cite{zhu2024malicious} can lead to catastrophic failures, such as misdetecting obstacles \cite{zhu2024malicious, sun2021control}, ignoring traffic signs \cite{jia2022fooling}, or miscalculating trajectories \cite{chen2024adversary}. These attacks endanger passenger safety and undermine public trust in autonomous technologies \cite{bathla2022autonomous}. Although research on sensor-specific attacks, such as LiDAR and radar spoofing and camera adversarial examples, has gained attention, a comprehensive evaluation of existing single-sensor and MSF perception attacks in dynamic real-world environments remains lacking.

Recent studies have demonstrated the feasibility of single-sensor attacks, such as LiDAR or radar spoofing to create ``phantom'' obstacles, and adversarial stickers to mislead camera-based object detection \cite{jin2024phantomlidar}. These attacks exposed vulnerabilities in individual sensors, highlighting the limitations of relying on single perception systems. In response, modern AVs are increasingly adopting cross-modal or MSF systems (e.g., camera, LiDAR, radar) to enhance perception accuracy and robustness by combining multiple sensor data. However, this shift toward MSF has introduced new attack surfaces, as adversaries can now target and expose inconsistencies between fused sensor outputs to hamper the system's reliability. For example, Zhu et al. \cite{zhu2024malicious} demonstrated that adversarial objects interfering with LiDAR, camera, and radar simultaneously can bypass redundancy checks, while Hallyburton et al. \cite{hallyburton2022security} introduced stealthy ``frustum attacks'' that maintain semantic consistency across camera and LiDAR modalities, making them harder to detect. These works have been crucial in exposing the vulnerabilities of AV perception. However, they lead to a pivotal, unanswered question: is the current research landscape complete and systematic enough to provide a comprehensive understanding of the threat model against modern AV perception, especially as the industry moves from single-sensor reliance to robust MSF?

Unfortunately, the answer is no. Despite prior research efforts, the field still lacks a comprehensive analysis of how attacks have evolved from targeting single sensors to exploiting fusion-based systems. The existing literature on AV perception attacks is fragmented. This fragmentation makes it challenging for researchers and practitioners to grasp the comprehensive adversarial landscape and understand the unique vulnerabilities introduced by MSF systems. Key questions remain: Have defenses adapted to these evolving attacks? What new attack methods remain unexplored? Where do critical vulnerabilities persist in MSF-based perception systems?

To bridge this gap, this paper presents a comprehensive systematization of knowledge (SoK) of attacks on AV perception systems with three key contributions:

\textbf{First}, we perform a comprehensive analysis and systematization of existing perception attacks based on 7 dimensions and reveal a fragmented and imbalanced research landscape through a systematic analysis of 48 foundational studies. We demonstrate that existing work is overwhelmingly biased toward single-sensor attacks (e.g., on cameras or LiDAR in isolation), a focus that masks the true, systemic vulnerabilities of modern MSF architectures.

\textbf{Second}, we establish the first unified taxonomy of 20 distinct perception attack methods, deconstructing the current attack spectrum. This taxonomy, which connects attack mediums (e.g., physical objects, signal spoofing) to their specific target modules (e.g., Traffic Light Recognition, Object Detection, Classification, Depth Estimation, Point Cloud Generation), creates a common framework for threat modeling that has been absent in the fragmented literature.

\textbf{Third}, we validate the criticality of this unexplored MSF threat landscape by proposing and digitally simulating a novel, cross-modal attack vector - the Combined IR Laser \& LiDAR Spoofing Attack to demonstrate the feasibility of defeating modern fusion logic. Our results prove that by systematically coordinating attacks across sensors, an adversary can create high-confidence ``phantom'' objects that are invisible to single-sensor defenses, confirming the gap identified by our systematization is a practical and significant threat.
\section{Methodology}
\label{sec:method}

To systematically analyze perception attacks targeting single-sensor and MSF systems in AVs, we adopted the structured five-step literature review process from Wolfswinkel et al. \cite{wolfswinkel2013using}: Define, Search, Select, Analyze, and Present.

\noindent\textbf{Define:} We defined the study scope by:
\begin{itemize}
    \item \textit{Sources:} Collecting peer-reviewed papers from security (e.g., IEEE S\&P, NDSS, USENIX Security, CCS), computer vision (e.g., CVPR, ICCV), robotics (e.g., CoRL, ICRA), and mobile/networking (e.g., MobiCom, SenSys) venues via major academic databases (Google Scholar, IEEE Xplore, ACM DL, etc.).
    \item \textit{Search Terms:} Using keywords including ``autonomous vehicles'', ``AV attacks'', ``multi-sensor fusion'', ``LiDAR spoofing'', ``camera spoofing'', ``radar spoofing'', ``AV perception attacks'', ``adversarial attacks'', and ``sensor security''.
    \item \textit{Inclusion Criteria:} Focusing on papers investigating attacks on AV perception (single-sensor or MSF), proposing relevant defenses, or analyzing sensor manipulation impact.
\end{itemize}

\noindent\textbf{Search:} We conducted an iterative search using these terms and sources, combined with backward \cite{jalali2012systematic} and forward \cite{felizardo2018evaluating} snowballing, yielded 124 initial candidate papers.

\noindent\textbf{Select:} We applied a three-stage filtering: (i) Abstract/Introduction screening to exclude papers outside the AV perception attack scope; (ii) We retained papers published between 2019--2025 to get state-of-the-art in perception attacks; (iii) Full-text review confirmed relevance to attack methodologies (single-sensor or MSF) and potential defenses. This resulted in a final corpus of 48 papers relevant to this SoK.

\noindent\textbf{Analyze:} We analyzed the final selected papers along two primary categories:
\begin{itemize}
    \item \textit{Attack Targets \& Evaluation Environments:} We characterized attacks by sensor modality (single vs. MSF) and evaluation settings (e.g., simulation through CARLA, datasets like KITTI, real-world tests with varying weather conditions etc.).
    \item \textit{Attack Types and Methods:} We identified several attack types (spoofing, adversarial examples, DoS, poisoning) and classified their specific methods. This analysis revealed publication trends over the past seven years, showing increased focus on LiDAR and radar alongside camera. MSF attack research, while currently less frequent, shows growing interest.
\end{itemize}
\noindent\textbf{Present:} The synthesized findings, including the taxonomy of attacks and methods, trends, and vulnerabilities, are presented in Section~\ref{sec:current-work}. This systematic approach ensures a comprehensive analysis of the current research landscape in AV perception attacks.
\section{Background and Technical Foundations}
\label{sec:background}
 
AV automation progresses through six levels from level 0 to level 5 \cite{khan2022level,zhang2022finding}. With each level progression, AVs' cyber-physical system becomes more complex, whose operational integrity depends on a pipeline of trusted data processing. This pipeline transforms raw sensor measurements into physical control commands through a sequence of interdependent modules. A breakdown in this chain, particularly at the perception stage, can lead to catastrophic system failures. This section formalizes the core components of the AV stack to establish the technical foundation for analyzing security vulnerabilities in subsequent sections.

\subsection{The AV Operational Pipeline}
\label{subsec:pipeline}

The standard AV operational pipeline, as illustrated in Figure~\ref{fig:attack-method}, comprises three critical modules \cite{pendleton2017perception,dai2020perception, fan2023autonomous, lee2024efficient, zhu2024malicious}: \textbf{1) Perception:} Processes raw sensor data (e.g., LiDAR point clouds, camera images, radar signals) to construct a semantic understanding of the environment, including object detection, classification, and tracking; \textbf{2) Planning:} Utilizes the perceived state to perform localization, mapping, and trajectory prediction, formulating a safe navigational policy for the AV; \textbf{3) Control:} Executes the planned policy via Electronic Control Units (ECUs) such as the Navigation Control Module (NCM) and Engine Control Module (ECM) by issuing steering, acceleration, and braking commands \cite{al2019autonomous, chow2021attack}. This pipeline operates under a critical trust assumption: each module relies on the correctness of its predecessor's output. Consequently, a perception error can cascade, causing incorrect planning decisions (e.g., faulty trajectory prediction \cite{li2021fooling}) and hazardous control actions (e.g., sudden braking or collisions \cite{chernikova2019self}).

\subsection{Perception System Architectures}
\label{subsec:perception}

The perception system is responsible for interpreting raw sensor data to infer the state of the AV and its environment. Its architecture can be broadly categorized into single-sensor and MSF paradigms.

\subsubsection{Single-Sensor Perception}
\label{subsubsec:single_sensor}

Single-sensor systems rely on one primary modality, making them susceptible to attacks that exploit the inherent physical limitations of that sensor.

\textbf{Camera Perception.} Camera sensors capture a 2D image $I_t$. A perception algorithm, typically a Deep Neural Network (DNN) like YOLO \cite{bochkovskiy2020yolov4} or Faster R-CNN \cite{ren2016faster}, processes this image to produce a set of 2D detections $D_C = \{(b_C, \text{class}_C, s_C)\}$, where $b_C$ is a bounding box, $\text{class}_C$ is an object label, and $s_C$ is a confidence score. In AV development, three primary camera types are utilized: monocular (single-lens), stereo (dual-lens), and specialized integrated cameras \cite{herman2017single}. The camera-only perception system constructs 3D representations of the environment by extracting features from multiple camera images and projecting them onto a 2D ground plane. This technique improves spatial understanding of the vehicle's surroundings \cite{harley_2023_simplebev}. These systems are cost-effective and provide high-resolution semantic understanding, enabling critical ADAS features like forward collision warning \cite{lim2018real}, traffic sign recognition \cite{ikhlayel2020traffic, alhabshee2020deep}, pedestrian recognition \cite{li2022detection, huang2021bevdet}, lane detection \cite{zakaria2023lane, gajjar2023comprehensive, singal2023roadway}, blind spot detection \cite{guo2018blind, bagi2019sensing} and automated parking assistance \cite{khalid2021smart}. Cameras are inexpensive compared to LiDAR and radar and thus are the preferred sensing modality for many AVs \cite{kato2018autoware,hecht2018lidar, hallyburton2022security}. However, they are vulnerable to adversarial perturbations \cite{jia2022fooling} and perform poorly in low-light conditions. 

\textbf{LiDAR Perception.} LiDAR is an active sensor that emits infrared laser pulses to generate a 3D point cloud $PC_t = \{\mathbf{p}_j \in \mathbb{R}^3\}_{j=1}^{M}$, where $\mathbf{p}_j = (x_j, y_j, z_j)$ represents the 3D Cartesian coordinates of the $j$-th point and $M$ is the total number of points in the cloud map. Algorithms like PointRCNN~\cite{shi2019pointrcnn} or PointPillar~\cite{lang2019pointpillars} process $PC_t$ to produce 3D object detections $D_L = \{(b_L, s_L)\}$. These systems use only laser-based data to detect and interpret objects in the environment, without the assistance of other sensors such as cameras or radar \cite{rigoulet_2022_lidar}. LiDAR provides precise depth information and a 360-degree field of view \cite{zhu20123d}, making it invaluable for 3D object detection and mapping. Its active nature, however, makes it susceptible to signal spoofing attacks that inject phantom points \cite{jin2024phantomlidar}.

\textbf{Radar Perception.} Automotive radar uses radio waves to measure the range, Doppler velocity, and azimuth angle of objects \cite{bilik2019rise, kocic2018sensors} by emitting radio waves toward target areas \cite{patole2017automotive, xu2023sok, kong2024survey}. Radar sensors operate on Frequency-Modulated Continuous Wave (FMCW) principles, transmitting chirp signals $s_{tx}(t)$ and processing the reflected signals $s_{rx}(t)$ to extract target information. The beat signal $s_{beat}(t) = s_{tx}(t) \cdot s_{rx}(t)$ enables simultaneous range and velocity estimation through Fast Fourier Transform (FFT) processing. It provides robust sensing in adverse weather conditions like fog and rain \cite{yao2023radar} but suffers from lower spatial and angular resolution compared to cameras and LiDAR \cite{yao2009458, li2022exploiting}. Radar data, which comes in formats like ADC signals, range-Doppler maps, and point clouds, is crucial for velocity estimation and object tracking. Emerging 4D mmWave radar technology enhances perception by adding elevation measurement \cite{fan20244d}.

\subsubsection{MSF Perception}
\label{subsec:fusion}

To overcome the limitations of individual sensors, AVs employ MSF architectures that combine data from complementary modalities like cameras, LiDAR, and radar \cite{bhupathiraju2023emi,zhu2021adversarial, cheng2023fusion, yang2022generating, zheng2024physical, cheng2022physical, jin2024phantomlidar, lou2024first, jin2023pla, liu2023slowlidar}. The two primary fusion strategies are:

\textbf{Early Fusion (Data-Level).}
This approach combines raw or low-level sensor data (e.g., projecting LiDAR point clouds onto camera images using calibration matrix $\mathbf{T}_{lidar}^{cam}$) before feature extraction \cite{kim2018robust, lim2019radar}. The fused features $\mathbf{F}_{early} = [\phi_{img}(I); \phi_{pc}(\mathbf{T}_{lidar}^{cam} \cdot PC)]$, where $\phi_{img}$ and $\phi_{pc}$ are feature extractors, $[\cdot]$ denotes concatenation, and $\mathbf{T}_{lidar}^{cam}$ the calibration matrix. The fusion must satisfy spatial consistency constraints. This enables joint learning but requires significant computational resources and precise sensor calibration.

\textbf{Late Fusion (Decision-Level).}
This method merges processed outputs from independent perception stacks (e.g., associating LiDAR bounding boxes $b_L$ with camera detections $b_C$ when $\texttt{IoU}(\texttt{Proj}(b_L), b_C) > \tau$), where $\texttt{Proj}()$ projects 3D boxes to image coordinates \cite{jahn2024enhancing, kuhn2020introspective}. It is more modular but vulnerable to attacks that deceive multiple sensors independently. While MSF is designed to enhance robustness, the fusion logic effectively becomes a new dependency; adversaries can target the specific algorithms used to reconcile conflicting sensor data, turning the redundancy mechanism into a point of failure~\cite{zhu2024malicious, hallyburton2022security}.

\subsection{Planning and Localization}
\label{subsec:planning}

The planning module uses the perceived environmental state for mapping, localization, and trajectory prediction. Localization estimates the vehicle's precise pose (often to centimeter-level accuracy) by fusing data from sensors like GPS and IMU, often using estimators like the Error-State Kalman Filter (ESKF) \cite{zhang2022play, han2024visionguard}. GPS provides absolute positioning \cite{girbes2021asynchronous}, while the IMU tracks acceleration and orientation, enabling dead reckoning when GPS is unavailable \cite{mousa2017inertial}. Mapping, which can be performed using services like HD maps or via Simultaneous Localization and Mapping (SLAM)~\cite{gao2020stereo}, provides the reference data for localization. Trajectory prediction forecasts the future motions of other agents using models like the Autoregressive Integrated Moving Average (ARIMA) to analyze time-series data \cite{alzyout2020performance, siami2018forecasting}. This module is vulnerable to spoofing attacks that corrupt sensor inputs, such as GPS spoofing \cite{shen2020drift} or LiDAR-based trajectory manipulation \cite{li2021fooling}, leading to severe navigation errors.
\section{Systematization Scope}

Earlier works on autonomous vehicle (AV) security have approached perception-layer vulnerabilities from fragmented perspectives, creating significant gaps in the current literature. We categorize these limitations into three main areas:

\textbf{Sensor-Specific and Attack-Limited Analyses.} Most prior works focus on isolated aspects of perception security. Some works concentrate on general security across robotic platforms \cite{altawy2016security, luo2016drones, nassi2021sok, li2020survey, petit2015remote, parkinson2017cyber, ren2019security, shen2022sok}, while others target specific sensor types such as GPS spoofing \cite{psiaki2016gnss, schmidt2016survey}, camera attacks \cite{huang2020survey}, or LiDAR vulnerabilities \cite{guesmi2024navigating, chi2024adversarial, kim2024survey}. Xu et al. \cite{xu2023sok} extended this by exploring various sensor spoofing attacks across different robotic vehicles, and Yan et al. \cite{yan2020sok} introduced sensor security models focusing on signal processing mechanisms. However, these works remain constrained to specific attack classes or sensor modalities, preventing a unified understanding of the interconnected threat landscape.

\textbf{Algorithm-Centric Security Perspectives.} Another stream of research focuses exclusively on machine learning vulnerabilities. Mahima et al. \cite{mahima2024toward} provided a comprehensive taxonomy of adversarial attacks on 3D perception tasks, but their analysis is confined to learning-based algorithms, overlooking physical-world attack vectors and signal-level manipulations.

\textbf{Fusion as Defense Rather Than Attack Surface.} Recent work by Xiang et al. \cite{xiang2023multi} presented taxonomies for MSF and cooperative perception, but treated MSF primarily as a robustness mechanism. Their analysis overlooks how fusion dependencies themselves create new attack vectors that can bypass single-sensor defenses. This SoK bridges these critical gaps by providing the first unified analysis that spans both single-sensor and MSF systems while covering the full spectrum of perception attacks. We deconstruct the threat landscape into a unified framework, categorizing distinct attack methods across the full spectrum of sensor modalities. We trace the progression from single-sensor exploitation to sophisticated cross-modal attacks and systematically reveal how multi-sensor fusion can introduce new attack surfaces rather than just providing robustness.

\textbf{Scope (AV vs. Broader Robotics).} In this SoK, we focused strictly on road AVs because they operate in a dynamic road environment with unique safety-critical constraints (e.g., high speeds, human passengers, mixed-traffic environments) and rely on a relatively standardized, complex MSF architecture (Camera + LiDAR + Radar). General robots and UxVs (e.g., drones) often operate in different domains (e.g., industry, household, 3D airspace) with vastly different, sometimes simpler, perception and localization stacks (e.g., heavily relying on IMU/GPS or purely optical flow). Expanding the scope would weaken the targeted analysis of automotive MSF vulnerabilities.

\section{Systematization of Knowledge of Existing AV Perception Attacks}
\label{sec:current-work}
This section systematically categorizes perception attacks targeting AV security, focusing on single-sensor and MSF vulnerabilities. The analysis is three-fold. In the first part (Section~\ref{subsec:5.1}), we systematize 48 studies on various attack vectors in AV perception by sensor type (e.g., camera, LiDAR, radar), attack stage, attack goals (e.g., misclassification, misdetection, trajectory perturbation), adversary knowledge level and evaluation environment of each study. We then further analyze and categorize each work by its targeted AV modules, identifying 20 attack methods across different attack media and classifying them into six attack classes (Section~\ref{subsec:5.2}). Finally, we present a comprehensive taxonomy of single-sensor perception attacks based on previous works and our analysis (Section~\ref{subsec:5.3}).

\begin{table*}[hbt!]
\centering
\small
\rowcolors{2}{gray!15}{white}
\begin{adjustbox}{width=\linewidth}
\begin{tabular}{|p{0.18\linewidth}|p{0.01\linewidth}|>{\centering\arraybackslash}p{0.005\linewidth}>{\centering\arraybackslash}p{0.005\linewidth}>{\centering\arraybackslash}p{0.005\linewidth}>{\centering\arraybackslash}p{0.01\linewidth}|>{\centering\arraybackslash}p{0.03\linewidth}|p{0.015\linewidth}p{0.015\linewidth}p{0.015\linewidth}|>{\centering\arraybackslash}p{0.025\linewidth}|>{\centering\arraybackslash}p{0.2\linewidth}|>{\centering\arraybackslash}p{.1\linewidth}|>{\centering\arraybackslash}p{.03\linewidth}|>{\centering\arraybackslash}p{.01\linewidth}|>
{\centering\arraybackslash}p{.01\linewidth}|}
\hline
\multirow{2}{*}{} & \multirow{2}{*}{} & \multicolumn{4}{c|}{\textbf{Attack Stage}} & \multirow{2}{*}{} & \multicolumn{3}{c|}{\textbf{Attack Goal}} & \multirow{2}{*}{} & \multirow{2}{*}{} & \multirow{2}{*}{} & \multirow{2}{*}{} & \multirow{2}{*}{} & \multirow{2}{*}{}\\
\cline{3-6} \cline{8-10}
\rowcolor{white}
\textbf{Reference} & \rotatebox{90}{\textbf{Perception Type}} & 
\rotatebox{90}{\textbf{Input-level}} & 
\rotatebox{90}{\textbf{Processing-level}} & 
\rotatebox{90}{\textbf{Early Fusion}} & 
\rotatebox{90}{\textbf{Late Fusion}} & 
\rotatebox{90}{\textbf{Target Attack Sensor}}& 
\rotatebox{90}{\textbf{Object Misclassification}} & 
\rotatebox{90}{\textbf{Misdetection \& Tracking}} & 
\rotatebox{90}{\textbf{Lane Misdetection}} &
\rotatebox{90}{\textbf{Simulation}} &
\textbf{Digital World (Dataset)} &
\rotatebox{90}{\textbf{Real World (Conditions)}} &
\rotatebox{90}{\textbf{Attack Surface}} & 
\rotatebox{90}{\textbf{Attacker's Knowledge}} & 
\rotatebox{90}{\textbf{MSF as Proposed Defense}}\\ \hline
Wang et al. \cite{wang2024revisiting}&  &  \checkmark &   &   &   & 1  &  \checkmark & & & \scross & COCO,ARTS & \twemoji{sunrise} \twemoji{sun} & P &  \Circle & \scross
 \\
Sato et al. \cite{sato2024invisible} &  &  \checkmark &   &   &   & 1 &  \checkmark &   & &  \scross & COCO,ARTS,LISA& \twemoji{sunrise} \twemoji{night with stars} \twemoji{sun} & P &  \Circle & \scross
 \\
Ma et al. \cite{ma2024slowtrack} &  &   &  \checkmark &   &   & 1  &   &  \checkmark & & 2 & MOT17,BDD100K& ? & P & \Circle & \scross
 \\
Zheng et al. \cite{zheng2024pi} &  &  \checkmark &   &   &   & 1  &   &  \checkmark & &  \scross & KITTI,KAIST & \twemoji{sunrise} \twemoji{sun} & P & \CIRCLE & \checkmark
 \\
Chen et al. \cite{chen2024adversary} &  &  \checkmark &   &   &   & 1   &  \checkmark &   & & \scross& KITTI & ? & P &  \LEFTcircle & \scross
 \\
Zheng et al. \cite{zheng2024physical}&  &  \checkmark &   &   &   & 1   &  \checkmark &   & & 1 & \scross& ? & P&    \CIRCLE & \scross
 \\
Yan et al. \cite{yan2022rolling} &  &  \checkmark &   &   &   & 1   &  \checkmark &  \checkmark & & 2,3 & BDD100K & \twemoji{sunrise} \twemoji{sun} & S   &  \CIRCLE & \scross
 \\
Jia et al. \cite{jia2022fooling}&  &  \checkmark &   &   &   & 1   &  \checkmark &  \checkmark & & \scross & TT100K & \twemoji{sunrise} \twemoji{night with stars} \twemoji{sun} \twemoji{cloud} & P &  \CIRCLE & \scross
 \\
Han et al. \cite{han2022physical} &  &   &  \checkmark &   &   & 1   &   &   &  \checkmark  &  \scross &TuSimple & \twemoji{sunrise} \twemoji{sun} & P &  \CIRCLE & \scross
 \\
Muller et al. \cite{muller2022physical} &  &  \checkmark &   &   &   & 1   &   &  \checkmark & &1 &VOT & \twemoji{sunrise} \twemoji{sun} \twemoji{cloud}& P &  \CIRCLE & \scross
 \\
Cheng et al. \cite{cheng2022physical}&  &  \checkmark &   &   &   & 1   &   &  \checkmark & & \scross&KITTI& \twemoji{sunrise} \twemoji{city_sunset}\twemoji{sun} \twemoji{cloud} & P&    \CIRCLE & \scross
 \\
Zhao et al. \cite{zhao2019seeing} &  &  \checkmark &   &   &   & 1   &  \checkmark &  \checkmark & &\scross& COCO& \twemoji{sun} \twemoji{cloud} & P &  \LEFTcircle & \scross
 \\
Man et al. \cite{man2020ghostimage} &  &  \checkmark &   &   &   & 1  &  \checkmark &   & &\scross&LISA& \twemoji{city_sunset} \twemoji{night with stars} \twemoji{sun} & S &   \CIRCLE & \scross
 \\
Ji et al. \cite{ji2021poltergeist} &  &   &  \checkmark &   &   & 1   &  \checkmark &  \checkmark & &\scross&KITTI,BDD100K& \twemoji{city_sunset} \twemoji{night with stars} \twemoji{sun} \twemoji{cloud with rain} \twemoji{fog} & S &    \CIRCLE & \checkmark
 \\
Sato et al. \cite{sato2021dirty}&  &  \checkmark &   &   &   & 1   &  \checkmark &   &  \checkmark &4& comma2k19& \twemoji{city_sunset}& P &    \CIRCLE & \checkmark
 \\
Jing et al. \cite{jing2021too} &  &  \checkmark &   &   &   & 1   &  \checkmark &   &  \checkmark &5&\scross&  \twemoji{city_sunset} \twemoji{night with stars}& P &    \CIRCLE & \checkmark
 \\
Wang et al. \cite{wang2021can} & \multirow{-17}{*}{\rotatebox{90}{Camera}} &  \checkmark &   &   &   & 1   &  \checkmark &  \checkmark & &\scross& KITTI,Bosch Night& \twemoji{sun} & P &   \CIRCLE & \checkmark
 \\ \hline
Jin et al. \cite{jin2024phantomlidar} &  &  \checkmark &   &   &   &2   &  \checkmark &  \checkmark &  &\scross& KITTI& \twemoji{sunrise} \twemoji{sun}&  S& \CIRCLE & \checkmark
 \\
Zhang et al. \cite{zhang2024online}&  &  \checkmark &   &   &   & 2   &  \checkmark &  \checkmark & &1& nuScenes & \twemoji{sunrise} \twemoji{sun} \twemoji{cloud} & P &    \CIRCLE & \scross
 \\
Lou et al. \cite{lou2024first} &  &   &  \checkmark &   &   & 2  &  \checkmark &  \checkmark & &\scross&KITTI&  \twemoji{sunrise} \twemoji{sun} & P &   \CIRCLE & \scross
 \\
Zhu et al. \cite{zhu2024ae}&  &   &  \checkmark &   &   & 2   &  \checkmark &  \checkmark & &1& KITTI,Apollo& \twemoji{sunrise} \twemoji{night with stars} \twemoji{sun}& P&  \Circle & \scross
 \\
Sato et al. \cite{sato2023lidar} &  &  \checkmark &   &   &   & 2   &  \checkmark &  \checkmark & &1& KITTI&  ? & S  & \Circle & \scross
 \\
Jin et al. \cite{jin2023pla}&  &  \checkmark &  &   &   & 2   &  \checkmark &  \checkmark & &\scross&KITTI& \twemoji{sunrise} \twemoji{night with stars} \twemoji{sun} \twemoji{cloud} & S  & \Circle & \checkmark
 \\
Liu et al. \cite{liu2023slowlidar} &  &  \checkmark &   &   &   & 2   &  \checkmark &  \checkmark & &\scross&KITTI& \twemoji{sunrise} \twemoji{sun} &  S &  \Circle & \scross
 \\
Bhupathiraju et al. \cite{bhupathiraju2023emi} &  &  \checkmark &   &   &   & 2  &  \checkmark &  \checkmark & &\scross&KITTI&  \twemoji{cloud} & S  &  \Circle & \scross
 \\
Cao et al. \cite{cao2023you} &  &  \checkmark &   &   &   & 2  &   &  \checkmark & &1& KITTI& \twemoji{sunrise} \twemoji{sun} \twemoji{cloud with rain} &  S &  \Circle & \scross
 \\
Li et al. \cite{li2023badlidet}&  &  \checkmark &   &   &   & 2   &   &  \checkmark & &1& SemanticKITTI&  ? & P &  \Circle & \checkmark
 \\
Li et al. \cite{li2023towards} &  &  \checkmark &   &   &   & 2 &   &  \checkmark & &1& KITTI& ? & P &  \Circle & \checkmark
 \\
Yang et al. \cite{yang2022generating} &  &  \checkmark &   &   &   & 2   &  \checkmark &  \checkmark & &\scross&nuScenes& ? & S  &  \Circle & \scross
 \\
Li et al. \cite{li2021fooling} &  &  \checkmark &   &   &   & 2   &  \checkmark &  \checkmark & &\scross& SemanticKITTI& ? & S  &  \Circle & \scross
 \\
Zhu et al. \cite{zhu2021adversarial} &  &  \checkmark &   &   &   & 2   &  \checkmark &  \checkmark &  &\scross&KITTI& \twemoji{sunrise} \twemoji{sun}& P &  \LEFTcircle & \checkmark
 \\
Zhu et al. \cite{zhu2021can} &  &  \checkmark &   &   &   & 2   &  \checkmark &  \checkmark & &\scross& KITTI& \twemoji{sunrise} \twemoji{sun} \twemoji{cloud} & P &   \CIRCLE & \scross
 \\
Tu et al. \cite{tu2020physically}&  &  \checkmark &   &   &   & 2   &  \checkmark &  \checkmark & &\scross& KITTI,Apollo& ?&P &  \LEFTcircle & \scross
 \\
Cao et al. \cite{cao2019adversarial} & \multirow{-17}{*}{\rotatebox{90}{LiDAR}} &  \checkmark &   &   &   & 2   &  \checkmark  &  \checkmark & &3&\scross&? &  S & \Circle & \checkmark
 \\ \hline
Sun et al. \cite{sun2021control}&  &  \checkmark &   &   &   & 3  &  &  \checkmark & & 3 & \scross& \twemoji{sunrise} \twemoji{sun} & S &  \Circle & \scross
 \\ 
Hadad et al. \cite{hadad2025adversarial} &  &  \checkmark &  &  &   & 3  &  \checkmark & & &\scross& Custom Radar Dataset & \twemoji{sunrise} \twemoji{sun} & S & \Circle & \scross
 \\
Komissarov et al. \cite{komissarov2021spoofing} &  &  \checkmark & & &   & 3  & &  \checkmark & & 6 &\scross& ? &S  & \Circle & \checkmark
 \\ 
Nallabolu et al. \cite{nallabolu2021frequency} &  &  \checkmark & & &   & 3  & &  \checkmark & & 7,8 &\scross& \twemoji{sunrise} \twemoji{sun} &S  & \Circle & \scross
 \\ 
Zhu et al. \cite{zhu2023tilemask} &  &  \checkmark & & &   & 3  &  \checkmark &  \checkmark & & \scross& SemanticKITTI& \twemoji{sunrise} \twemoji{sun} &P  & \LEFTcircle & \checkmark
\\
Chen et al. \cite{chen2023metawave} &  &  \checkmark & & &   & 3  &  \checkmark &  \checkmark & & \scross& \scross& \twemoji{sun} \twemoji{night with stars} \twemoji{fog} \twemoji{2744}&P  & \CIRCLE & \scross
\\
Hunt et al. \cite{hunt2023madradar} & \multirow{-7}{*}{\rotatebox{90}{Radar}} &  \checkmark & & &   & 3  &  \checkmark &  \checkmark & & 6,7 & \scross& \twemoji{sunrise} \twemoji{sun}&S  & \CIRCLE & \checkmark
\\
\hline
Zhu et al. \cite{zhu2024malicious} &  &  \checkmark &   &  \checkmark &   & 1,2,3   &  \checkmark &  \checkmark & &3& SemanticKITTI& \twemoji{sunrise} \twemoji{sun} & P  & \Circle & \scross
 \\
Cheng et al. \cite{cheng2023fusion} &  &  \checkmark &   &  \checkmark &   & 1   &  \checkmark &  \checkmark & &\scross&KITTI& ?& P  & \Circle & \checkmark
 \\
Hallyburton et al. \cite{hallyburton2022security} &  &   &  \checkmark &   &  \checkmark & 2 &  \checkmark &  \checkmark & &1& KAIST& \twemoji{sunrise} \twemoji{sun} & P  &    \CIRCLE & \scross
 \\
Zhang et al. \cite{zhang2022play}&  &   &  \checkmark &   &   & 2,4   &   & \checkmark  & & \scross& \scross& ? & S   &    \CIRCLE & \scross
 \\
Tu et al. \cite{tu2021exploring} &  &  \checkmark &   &   &  \checkmark & 1,2   &  \checkmark &  \checkmark & &\scross& KITTI,Xenith& ? & P & \LEFTcircle & \scross
 \\
Cao et al. \cite{cao2021invisible} &  &  \checkmark &   &  \checkmark &   & 1,2  &  \checkmark &  \checkmark & &2,3& KITTI& \twemoji{sunrise} \twemoji{sun} &P  & \Circle & \scross
 \\
Shen et al. \cite{shen2020drift} & \multirow{-7}{*}{\rotatebox{90}{MSF}} &  \checkmark &   &   &  \checkmark & 4 &   & \checkmark  & &\scross&KAIST& \twemoji{cloud with rain} \twemoji{fog} &  S  & \Circle & \checkmark \\ \hline
\end{tabular}
\end{adjustbox}
\begin{minipage}{17.5cm}
\vspace{0.1cm} 
\footnotesize \textbf{Target Attack Sensor:} 1 = Camera, 2 = LiDAR, 3 = Radar, 4 = GPS; \textbf{Simulation:} 1 = CARLA, 2 = LGSVL, 3 = Baidu Apollo, 4 = OpenPilot, 5 = AutoPilot, 6 = SDR, 7 = MATLAB, 8 = AWR VSS, \scross = Not Used; \textbf{Real World (Conditions):} \twemoji{sunrise}: Noon, \twemoji{city_sunset}: Dusk, \twemoji{night with stars}: Night, \twemoji{sun}: Sunny, \twemoji{cloud}: Cloudy, \twemoji{cloud with rain}: Rainy, \twemoji{fog}: Foggy, \twemoji{2744}: Snowy, Not Found: ?; \textbf{Attacker Knowledge}: $\Circle$ = White-Box, $\LEFTcircle$ = Gray-Box, $\CIRCLE$ = Black-Box
\end{minipage}
\caption{A summary of existing research on attacks targeting perception systems in autonomous vehicles, detailing the sensor type, attack stage, objectives, datasets, simulation tools, and attacker knowledge assumptions to provide a comparative view of how different studies evaluate the robustness of AV perception against various threats.}

\label{tab:attack}
\end{table*}

\subsection{Taxonomy of Attack Properties}
\label{subsec:5.1}
Our primary systematization, summarized in Table~\ref{tab:attack}, classifies the finalized foundational studies along 7 key dimensions. This analysis moves beyond simple enumeration to reveal structural patterns, biases, and the architectural implications of the current AV security landscape.

\subsubsection{Dimension 1: Perception Modality}

This dimension defines the primary axis of the taxonomy and classifies prior studies by the perception modality they attack. Table~\ref{tab:attack} shows a clear imbalance: most existing work focuses on single sensor vulnerabilities, with camera, LiDAR, and radar examined independently. Only a small number of studies investigate the MSF layer, even though it plays a central role in providing redundancy and improving robustness. This imbalance highlights a critical blind spot in current research, since the component designed to strengthen AV perception remains the least examined attack surface.

\begin{tcolorbox}[colback=gray!10, colframe=black, boxrule=0.4pt, arc=2pt, left=2pt, right=2pt, top=2pt, bottom=2pt]
\textbf{Insight 1: A Component-Level Validation Focus.} 

\textit{The concentration of attacks on individual sensors reveals that defenses have evolved in a modality-siloed manner. In practice, however, perception safety depends on the fusion logic that reconciles cross-modal signals. The sparse attention to MSF therefore exposes a structural security gap: the most safety-critical component receives the least adversarial scrutiny.}

\end{tcolorbox}

\subsubsection{Dimension 2: Attack Stage (Interference Point)}
We classify attacks based on \textit{when} they interfere with the AV perception pipeline: \textbf{Input-Level Attacks:} These manipulate raw sensor data before processing, such as deploying adversarial patches on camera images \cite{wang2024revisiting, chen2024adversary} or vehicles \cite{zheng2024physical, cheng2022physical}, placing adversarial objects \cite{zheng2024pi, jia2022fooling}, or spoofing LiDAR point clouds \cite{jin2024phantomlidar} and radar signals \cite{sun2021control, hunt2023madradar}, \textbf{Processing-Level Attacks:} These attacks target perception algorithms directly, such as slowing object tracking \cite{ma2024slowtrack} or poisoning training data \cite{han2022physical}, \textbf{Early and Late Fusion Attacks:} These exploit multi-sensor integration. Early fusion attacks manipulate data before combination \cite{chen2022cooperative, kumar2012carspeak}, while late fusion attacks alter outputs after fusion \cite{shi2022vips, song2023cooperative}.

\begin{tcolorbox}[colback=gray!10, colframe=black, boxrule=0.4pt, arc=2pt, left=2pt, right=2pt, top=2pt, bottom=2pt]
\textbf{Insight 2: Input-Level Dominance vs. Emerging Algorithmic Threats.} \textit{The corpus shows a strong bias toward input-level manipulation (41 of 48 studies), where altering raw sensor signals circumvents higher-layer protections. Processing-level (7 studies) and fusion-level (6 studies) attacks remain comparatively rare, marking the beginning of a shift toward deeper algorithmic and decision-logic exploitation.}
\end{tcolorbox}

\subsubsection{Dimension 3: Attack Goal (Consequence)}

This dimension in Table~\ref{tab:attack} defines the adversary's desired outcome. The attacker's objective typically falls into three main consequences: \textbf{Object Misclassification/Misinterpretation:} Causing the AV to assign an incorrect label to an object (e.g., misreading a stop sign \cite{wang2024revisiting, sato2024invisible}, or misclassifying a vehicle \cite{jin2024phantomlidar, bhupathiraju2023emi}), \textbf{Object Misdetection \& Tracking:} Causing the AV to fail to detect a present object (false negative) or to detect a \textit{Phantom} object (false positive \cite{ma2024slowtrack, muller2022physical, zhang2024online}), \textbf{Lane Misdetection:} Causing the AV to misinterpret road markings and drift out of its lane \cite{han2022physical, jing2021too}. 

\begin{tcolorbox}[colback=gray!10, colframe=black, boxrule=0.4pt, arc=2pt, left=2pt, right=2pt, top=2pt, bottom=2pt]
\textbf{Insight 3: Modality-Specific Tactical Objectives.} \textit{Camera, LiDAR, and radar attacks produce distinct semantic, geometric, and kinematic failures, respectively highlighting the need for fusion algorithms that detect cross-modal inconsistencies.}
\end{tcolorbox}

\subsubsection{Dimension 4: Attacker's Knowledge}
This dimension in Table~\ref{tab:attack} highlights a necessary trade-off between an attack's realism and its technical requirements. We adopt the standard knowledge models defined by Abdullah et al. \cite{abdullah2021sok}: \textbf{White-box ($\Circle$):} Attacker has full knowledge of the system, including model architecture, parameters, and gradients \cite{alpirez2019white, zheng2024pi, cao2023you, ma2024slowtrack, zhu2024ae, sato2023lidar}, \textbf{Black-box ($\CIRCLE$):} In this process, attacker has no internal knowledge; attacks rely on transferability or observable inputs/outputs \cite{sato2021dirty, jing2021too, wang2021can, man2020ghostimage}, \textbf{Gray-box ($\LEFTcircle$):} The attacker has partial knowledge, such as the model architecture but not the weights \cite{chen2024adversary, zhu2021adversarial, zhao2019seeing, tu2020physically, tu2021exploring}.

\begin{tcolorbox}[colback=gray!10, colframe=black, boxrule=0.4pt, arc=2pt, left=2pt, right=2pt, top=2pt, bottom=2pt]
\textbf{Insight 4: Black-box Feasibility for Visual Attacks.} \textit{Black-box attacks are largely confined to camera systems, while LiDAR and radar attacks rely on white-box assumptions. This asymmetry reflects the greater practicality of visual manipulation against camera-based perception.}
\end{tcolorbox}

\subsubsection{Dimension 5: Attack Surface}
Attacks are conducted via two primary surfaces: \textbf{Physical-World (P):} Manipulating physical objects or environments to exploit systems that rely on physical inputs \cite{man2020ghostimage,wang2024revisiting, sato2024invisible, ma2024slowtrack, zheng2024pi, zhu2023tilemask, chen2023metawave}, \textbf{Sensor (S):} Manipulating internal sensor data or components directly (e.g., via laser beams \cite{yan2022rolling}, EMI signals \cite{bhupathiraju2023emi}, acoustic signals \cite{ji2021poltergeist}, or RF signals \cite{komissarov2021spoofing}).

\begin{tcolorbox}[colback=gray!10, colframe=black, boxrule=0.4pt, arc=2pt, left=2pt, right=2pt, top=2pt, bottom=2pt]
\textbf{Insight 5: Correlation of Surface, Cost, and Knowledge.} \textit{Camera attacks cluster around physical-world manipulation, while LiDAR and radar attacks rely on sensor-level interference tied to white-box assumptions. Physical-world vectors, including most MSF attacks, therefore represent the more accessible and scalable threat class.}
\end{tcolorbox}

\subsubsection{Dimension 6: Evaluation Environment}
The choice of environment, \textbf{Simulation} (e.g., CARLA), \textbf{Digital-World} (static datasets like KITTI \cite{liao2022kitti}), or \textbf{Real-World} (physical testing), determines the ecological validity of a paper's claims and its transferability to the physical world. We analyzed the validation methods used in the prior studies in Table~\ref{tab:attack}: \textbf{Simulation:} Controlled, reproducible virtual environments (e.g., CARLA, LGSVL, Baidu Apollo), \textbf{Digital-World (Datasets):} Using pre-recorded real-world sensor data (e.g., KITTI \cite{liao2022kitti}, nuScenes \cite{caesar2020nuscenes}, COCO \cite{lin2014microsoft}), \textbf{Real-World:} Physical testing in practical driving conditions with varying weather and lighting.

\begin{tcolorbox}[colback=gray!10, colframe=black, boxrule=0.4pt, arc=2pt, left=2pt, right=2pt, top=2pt, bottom=2pt]
\textbf{Insight 6: A Pervasive `Evaluation Constraint'.} \textit{A majority of works assess attacks in digital or simulated environments, while only a subset validates them physically. While this reflects legitimate practical constraints, including cost, safety regulations, and equipment availability, this methodological gap limits understanding of environmental robustness and cross-domain transferability. Future works should address this through controlled test-track studies or hardware-in-the-loop approaches.}
\end{tcolorbox}

\subsubsection{Dimension 7: MSF as Proposed Defensive Approach}

This dimension in Table~\ref{tab:attack}'s final column analyzes a critical pattern in the literature: proposals that rely on MSF as a possible defense against a newly demonstrated single-sensor attack, effectively placing unverified trust in the integration layer.

\begin{tcolorbox}[colback=gray!10, colframe=black, boxrule=0.4pt, arc=2pt, left=4pt, right=4pt, top=4pt, bottom=4pt]
\textbf{Insight 7: The Escalating but Incomplete Reliance on Fusion.} \textit{The majority of studies ($\approx$75\%) either propose no defense or focus on narrow, attack-specific countermeasures. Only a small subset (16/48 papers) introduces defenses that are explicitly designed for multi-sensor or fusion-level threats. This reflects a reactive and fragmented defense landscape, where mitigation strategies lag behind the evolving attack vectors, especially in the context of MSF systems. Moreover, this defensive reliance is often unverified, as the papers provide no empirical evidence that the recommended MSF logic can handle the specific type of vulnerability discovered, especially against coordinated or fusion-aware attacks.}
\end{tcolorbox}

To illustrate the systematic bias in the research, we can see how the attack characteristics align: the less complex, easier-to-execute camera attacks often assume the Black-box setting and are validated using widely available Digital-World Datasets. Conversely, the more technically demanding LiDAR/Radar signal spoofing attacks require a White-box setting and often utilize the Sensor Attack Surface, highlighting the specialized knowledge and resources required to explore these high-impact threats.


\subsection{Analysis of Distinct Attack Methods Targeting AV Perception Modules}
\label{subsec:5.2}
We have analyzed and found the following important AV perception modules, such as Traffic Sign Recognition (TSR), Monocular Depth Estimation (MDE), vSLAM, Lane Detection, Object Detection \& Tracking, Classification, Point Cloud Generation, and Semantic Segmentation, are susceptible to adversarial manipulations. This subsection systematically categorizes adversarial methods into 3 attack mediums, 6 distinct attack classes, and 20 specific attack methods, as detailed in Table~\ref{tab:attack-mediums}.

\begin{table*}[hbt!]
\centering
\rowcolors{2}{gray!15}{white}
\begin{adjustbox}{width=\textwidth}
\begin{tabular}{|>{\centering\arraybackslash}p{0.14\linewidth}|>{\centering\arraybackslash}p{0.04\linewidth}|p{0.003\linewidth}p{0.003\linewidth}p{0.003\linewidth}p{0.003\linewidth}p{0.008\linewidth}|p{0.005\linewidth}p{0.005\linewidth}p{0.005\linewidth}p{0.005\linewidth}p{0.005\linewidth}p{0.005\linewidth}p{0.005\linewidth}p{0.008\linewidth}|p{0.005\linewidth}p{0.005\linewidth}p{0.005\linewidth}p{0.005\linewidth}p{0.005\linewidth}p{0.007\linewidth}|>{\centering\arraybackslash}p{0.06\linewidth}|>{\centering\arraybackslash}p{0.13\linewidth}|>{\centering\arraybackslash}p{0.09\linewidth}|}
\hline
\multirow{2}{*}{} & \multirow{2}{*}{} & \multicolumn{5}{c|}{\textbf{Adv. Patch}} & \multicolumn{8}{c|}{\textbf{3D Adversarial Object}} & \multicolumn{6}{c|}{\textbf{Signal Spoofer}} & \multirow{2}{*}{} & \multirow{2}{*}{} & \multirow{2}{*}{} \\
\cline{3-21}
\rowcolor{white}
\textbf{Targeted Module} & \rotatebox{90}{\textbf{Reference}} & \rotatebox{90}{3D textures} & \rotatebox{90}{Stickers} & \rotatebox{90}{Graffiti} & \rotatebox{90}{Image} & \rotatebox{90}{Color, Text, Shape}
 & \rotatebox{90}{Traffic Signs} & \rotatebox{90}{Traffic Cones} & \rotatebox{90}{Cardboard} & \rotatebox{90}{Rooftop Cargo} & \rotatebox{90}{Grass, Poles, Ladders} & \rotatebox{90}{Drones} 
 & \rotatebox{90}{Reflective Objects} & \rotatebox{90}{Random Object} & \rotatebox{90}{Noise} & \rotatebox{90}{Acoustic} & \rotatebox{90}{Laser} 
 & \rotatebox{90}{IEMI} & \rotatebox{90}{GNSS} & \rotatebox{90}{RF Signal} & \rotatebox{90}{\textbf{Attack Classification}} & \rotatebox{90}{\textbf{Attack Goal}} & \rotatebox{90}{\textbf{Attack Method}}\\ \hline

& \cite{wang2024revisiting} &  & \checkmark &  \checkmark &  &  &  \checkmark &  &  &  &  &  &  &  &  &  &  &  &  & & AP & HA& AtkMtd4\\ 

\cellcolor{white}&\cite{sato2024invisible}&  &  &  &  &  &   \checkmark &  &  &  &  &  &  &  &  &  &  \checkmark &  &  & & SS& NTA& AtkMtd5\\
 & \cite{yan2022rolling} &  &  &  &  &  &  &  &  &  &  &  &  &  &  &  &  \checkmark &  & &  & SS& TA& AtkMtd16\\ 
\cellcolor{white} \multirow{-4}{*}{TSR}& \cite{jia2022fooling}&  \checkmark &  &  \checkmark &  &  &  \checkmark &  &  &  &  &  &  &  &  &  &  &  &  & & PAO& HA, AA, TA, NTA& AtkMtd4\\ \hline
& \cite{ma2024slowtrack} &  &  &  &  &  &  &  &  &  &  &  &  &  &  \checkmark &  &  &  &  & & DoS& AA& AtkMtd9\\
\cellcolor{white}\multirow{-2}{*}{Processing Time}  & \cite{liu2023slowlidar} &  &  &  &  &  &  &  &  &  &  &  &  &  &  &  &  \checkmark &  & &  & DoS& HA, AA& AtkMtd10\\ \hline
&  \cite{zheng2024pi}&  &  &  &  &  &  \checkmark &  &  &  &  \checkmark &  &  &  &  &  &  &  & &  & PO& MIMO& AtkMtd2\\
\cellcolor{white}&  \cite{zheng2024physical} &  \checkmark &  &  &  &  &  &  &  &  &  &  &  &  &  &  &  &  & &  & AP& HA& AtkMtd6\\ 
\multirow{-3}{*}{MDE} & \cite{cheng2022physical}&  \checkmark &  &  &  &  &  &  &  &  &  &  &  &  &  &  &  &  & &  & AP& HA& AtkMtd6\\ \hline
\cellcolor{white}&  \cite{chen2024adversary}&  &  \checkmark &  &  \checkmark &  &  &  &  &  &  &  &  &  &  &  &  &  &  & & AP& HA& AtkMtd3\\
\multirow{-2}{*}{Visual SLAM} & \cite{wang2021can} &  &  &  &  &  &  &  &  &  &  &  &  &  &  &  &  \checkmark &  &  & & SS& HA& AtkMtd17\\ \hline
\cellcolor{white}&  \cite{han2022physical}&  &  &  &  &  &  &  \checkmark &  &  &  &  &  &  &  &  &  &  & &  & DP& MIMO& AtkMtd2\\
 &  \cite{sato2021dirty}&  &  \checkmark &  &  &  \checkmark &  &  &  &  &  &  &  &  &  &  &  &  &  & & AP& MIMO& AtkMtd3\\
\cellcolor{white}\multirow{-3}{*}{Lane Detection}  & \cite{jing2021too}&  &  &  &  \checkmark &  &  &  &  &  &  &  &  &  &  &  &  &  & &  & AP& MIMO& AtkMtd3\\ \hline
 &  \cite{muller2022physical}&  &  &  &  &  &  &  &  &  &  &  &  &  &  \checkmark &  &  &  & &  & SS& MIMO& AtkMtd9\\
\cellcolor{white} & \cite{zhao2019seeing}&  &  &  &  &  \checkmark &  &  &  &  &  &  &  &  &  &  &  &  & &  & AP& HA, AA& AtkMtd4\\
 & \cite{man2020ghostimage}&  &  &  &  &  &  &  &  &  &  &  &  &  &  &  &  \checkmark &  &  & & SS& HA, AA& AtkMtd5\\
\cellcolor{white} & \cite{ji2021poltergeist}&  &  &  &  &  &  &  &  &  &  &  &  &  &  &  \checkmark &  &  & &  & SS& HA, AA, NTA& AtkMtd9\\
 & \cite{wang2021can}&  &  &  &  &  &  &  &  &  &  &  &  &  &  &  &  \checkmark &  &  & & SS& HA& AtkMtd17\\
\cellcolor{white} & \cite{zhang2024online}&  &  &  &  &  &  &  &  &  &  &  &  &  \checkmark &  &  &  &  & &  & PAO& HA& AtkMtd7\\
 & \cite{li2023badlidet}&  &  &  &  &  &  \checkmark &  &  &  &  &  \checkmark &  &  &  &  &  &  &  & & DP& HA& AtkMtd14\\
\cellcolor{white} & \cite{cheng2023fusion}&  &  &  \checkmark &  &  \checkmark &  &  &  &  &  &  &  &  &  &  &  &  & &  & AP& HA& AtkMtd19\\ 
 & \cite{zhu2024malicious}&  &  &  &  &  \checkmark &  &  &  \checkmark &  &  &  \checkmark &  &  &  &  &  &  & &  & PAO& HA& AtkMtd12\\ 
\cellcolor{white}& \cite{sun2021control} &  &  &  &  &  &  &  &  &  &  &  &  &  &  &  &  &  & & \checkmark & SS& AA& AtkMtd13\\ 
 & \cite{hadad2025adversarial} &  &  &  &  &  &  &  &  &  &  &  &  &  &  &  &  &  & & \checkmark & SS& NTA& AtkMtd13\\
\cellcolor{white}& \cite{komissarov2021spoofing} &  &  &  &  &  &  &  & &  &  &  &  &  &  &  &  &  & & \checkmark & SS& HA& AtkMtd13\\
 & \cite{nallabolu2021frequency} &  &  &  &  &   &  &  &   &  &  &  &  &  &  &  &  &  & & \checkmark & SS& HA & AtkMtd13\\ 
\cellcolor{white} & \cite{zhu2023tilemask}&  &  &  &  &  &  &  &  &  &  &  & \checkmark &  &  &  &  &  &  & & AP & HA& AtkMtd12\\ 
\multirow{-15}{*}{\makecell{Object Detection, \\ Tracking, \& \\Classification}}& \cite{chen2023metawave}&  &  &  &  &  &  &  &  &  &  &  &  & \checkmark &  &  &  &  & &  & PAO& HA, AA& AtkMtd20\\ 
\cellcolor{white} & \cite{hunt2023madradar}&  &  &  &  &  &  &  &  &  &  &  &  &  &  &  &  &  & & \checkmark & SS& HA, AA& AtkMtd13\\ 
 \hline 
 & \cite{jin2024phantomlidar}&  &  &  &  &  &  &  &  &  &  &  &  &  &  &  &  &  \checkmark &  & & SS, DoS& HA, AA, NTA& AtkMtd18\\
\cellcolor{white} & \cite{zhu2024ae}&  &  &  &  &  &  &  &  &  &  &  &  &  \checkmark &  &  &  &  &  & & PAO& HA& AtkMtd20\\
 & \cite{sato2023lidar}&  &  &  &  &  &  &  &  &  &  &  &  &  &  &  &  \checkmark &  &  & & SS& HA, AA& AtkMtd18\\
\cellcolor{white} & \cite{jin2023pla}&  &  &  &  &  &  &  &  &  &  &  &  &  &  &  &  \checkmark &  & &  & SS& HA, AA& AtkMtd18\\
 & \cite{liu2023slowlidar}&  &  &  &  &  &  &  &  &  &  &  &  &  &  &  &  \checkmark &  &  & & DoS& HA, AA& AtkMtd18\\
\cellcolor{white} & \cite{bhupathiraju2023emi}&  &  &  &  &  &  &  &  &  &  &  &  &  &  &  &  &  \checkmark & &  & SS& HA, AA, NTA& AtkMtd18\\
 & \cite{cao2023you}&  &  &  &  &  &  &  &  &  &  &  &  &  &  &  &  \checkmark &  & &  & SS& HA& AtkMtd18\\
\cellcolor{white} & \cite{yang2022generating}&  &  &  &  &  &  &  &  &  &  &  &  &  &  &  &  &  &  & & SS& HA, AA& AtkMtd18\\
\multirow{-9}{*}{\makecell{Point Cloud \\ Generation}} & \cite{cao2019adversarial} &  &  &  &  &  &  &  &  &  &  &  &  &  &  &  &  \checkmark &  & &  & SS& AA& AtkMtd18\\ \hline
\cellcolor{white} & \cite{li2023towards}&  &  &  &  &  &  \checkmark &  &  &  &  &  \checkmark &  &  &  &  &  &  &  & & DP& HA& AtkMtd14\\
 & \cite{zhu2021adversarial}&  &  &  &  &  &  \checkmark &  &  \checkmark &  &  &  &  &  &  &  &  &  & &  & PO& HA& AtkMtd15\\
\cellcolor{white} & \cite{zhu2021can}&  &  &  &  &  &  &  &  &  &  &  \checkmark &  \checkmark &  &  &  &  &  &  & & AP& HA& AtkMtd8\\
 & \cite{tu2020physically}&  &  &  &  &  &  &  &  &  \checkmark &  &  &  &  &  &  &  &  &  & & PAO& HA& AtkMtd20\\
\cellcolor{white} & \cite{tu2021exploring}&  \checkmark &  &  &  &  &  &  &  &  &  &  &  &  &  &  &  &  &  & & PAO& HA& AtkMtd20\\
\multirow{-6}{*}{\makecell{Semantic \\Segmentation}} & \cite{cao2021invisible}&  &  &  &  &  &  &  \checkmark &  &  &  &  &  &  &  &  &  &  & &  & PAO& HA& AtkMtd2\\ \hline
\cellcolor{white} & \cite{lou2024first} &  &  &  &  &  &  &  &  \checkmark &  &  &  &  &  &  &  &  &  &  & & PO& MIMO& AtkMtd2\\
\multirow{-2}{*}{\makecell{Trajectory Prediction}} & \cite{li2021fooling}&  &  &  &  &  &  &  &  &  &  &  &  &  &  &  &  &  &  \checkmark & & SS& AA& AtkMtd11\\ \hline
 
\end{tabular}
\end{adjustbox}
\begin{minipage}{17.5cm}
\vspace{0.1cm}
\footnotesize \textbf{Attack Taxonomy:} AP = Adv. Perturbation, PAO = Physical Adv. Object, PO = Physical Object, DP = Data Poisoning, DoS = Denial of Service, SS = Signal Spoofing; HA = Hiding Attack, AA/CA = Appearing/Creating Attack, MIMO = Move-In, Move-Out Attack, TA = Target Attack, NTA = Non-Target Attack
\end{minipage}
\caption{A comprehensive overview of 20 attack vectors used against AV perception systems, categorized by target module, attack medium (e.g., adversarial patches, 3D objects, signal spoofers), classification type, and attacker objectives}
\label{tab:attack-mediums}
\end{table*}


\subsubsection{Attack Mediums used to conduct sensor attacks}
The attack mediums used in perception attacks are categorized in 3 classes:

\noindent\textbf{Adversarial Patches:} Adversarial patches exploit the vulnerabilities of AV vision-based systems by embedding deceptive patterns into the environment. These patches can take various forms, including:

\textit{3D Textures, Stickers, and Graffiti:} These perturbations alter the visual representation of objects, leading to misclassification of traffic signs and lane markings \cite{wang2024revisiting, jia2022fooling, sato2021dirty, cheng2023fusion}. \textit{Image Manipulation and Pattern-based Perturbations:} Image-based attacks, including altering the text, shape, or color of objects, have been found quite effective against object detection \cite{zhao2019seeing, cheng2023fusion} and lane detection \cite{sato2021dirty}.\medskip

\noindent\textbf{3D Adversarial Objects:}
Beyond 2D modifications, attackers can manipulate the physical world using adversarial objects that interfere with perception sensors. These include:

\textit{Traffic Signs, Cones, and Cardboard Objects:} Modified traffic signs and cones can cause TSR systems to misclassify speed limits or stop signs, while cardboard cutouts can create phantom obstacles for object detection \cite{wang2024revisiting, zheng2024pi, jia2022fooling}. Traffic cones can also be used to conduct backdoor attacks to hamper the lane detection module \cite{han2022physical}. \textit{Rooftop Cargo and Reflective Objects:} Attacks using cargo on vehicle rooftops introduce false reflections or shadows that impact AV detection \cite{muller2022physical, zhu2024ae}. Reflective materials such as metal-foil wrapped objects can deceive Radar-based depth estimation separately \cite{zhu2023tilemask} or simultaneously \cite{zhu2024malicious} along with LiDAR and camera sensors. \textit{Drones and Random Objects:} Unexpected moving objects such as drones or random objects (Meta-material tags, 3D-printed geometric shapes) can interfere with object detection and tracking, leading to false positives or occlusions \cite{zhu2024malicious, li2023badlidet, chen2023metawave, cao2021invisible}. \textit{Grass, Poles, and Ladders:} Natural objects are sometimes exploited to create false environmental features, tricking MDE module of AVs \cite{zheng2024pi}.\medskip

\noindent\textbf{Signal Spoofers:} Signal spoofing (SS) techniques exploit vulnerabilities in AV communication and sensor-based detection:

\textit{Noise Spoofing:} Some attacks target object tracking latency by increasing fake tracking boxes, causing AVs to experience delayed detection \cite{ma2024slowtrack, muller2022physical}. This is caused by inducing noise in the camera module. \textit{Acoustic and Laser Spoofing:} Acoustic and laser-based attacks can interfere with camera and LiDAR sensors by blurring perceived objects and/or injecting fake objects, introducing object misdetection, misclassification, false depth information or hallucinated obstacles \cite{muller2022physical, man2020ghostimage, ji2021poltergeist, wang2021can}. \textit{Intentional Electromagnetic Interference (IEMI) and GNSS Spoofing:} IEMI can disable or distort AV sensors \cite{jin2024phantomlidar, bhupathiraju2023emi, yang2022generating}, while GNSS spoofing affects localization and trajectory prediction, leading to navigation errors \cite{li2021fooling, shen2020drift}. \textit{Radio Frequency Signal:} RF signals are found to be used most frequently to disrupt the raw radar input in AV perception by generating false positives (e.g., phantom obstacles \cite{sun2021control}) or negatives (e.g., ``vanishing'' real vehicles \cite{hunt2023madradar}) by precisely manipulating chirp signals. It is also used for velocity spoofing to alter perceived object dynamics, causing AVs to misinterpret braking/acceleration patterns \cite{nallabolu2021frequency}.


\subsubsection{Discovered Attack Methods}
\begin{figure*}[hbt!]
    \centering
    \includegraphics[width=1\linewidth]{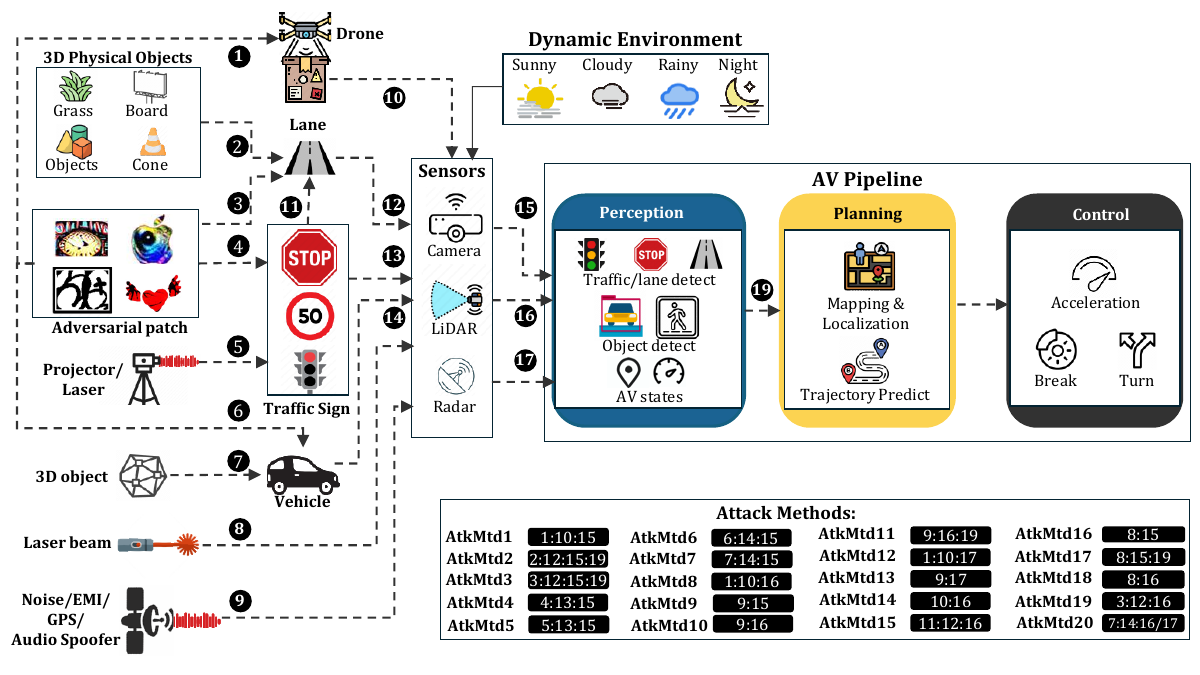}
    \caption{Overview of the Autonomous Vehicle (AV) system pipeline and a taxonomy of adversarial attack methods explored in prior research.}
    \label{fig:attack-method}
\end{figure*}

In this SoK work, we have discovered 20 attack methods (\textbf{AtkMtd1} to \textbf{AtkMtd20}) that can be used to attack the AV pipeline through various sensors, each targeting specific components of the AV stack. They are summarized in Figure~\ref{fig:attack-method}. These methods exploit various mediums explained previously and are designed to disrupt critical AV modules and pipeline.

It is found that \textbf{AtkMtd1} \tikz[baseline]{\node[anchor=base, draw, rounded corners] {1:10:15};}, \textbf{AtkMtd3} \tikz[baseline]{\node[anchor=base, draw, rounded corners]{3:12:15:19};}, \textbf{AtkMtd4} \tikz[baseline]{\node[anchor=base, draw, rounded corners] {4:13:15};}, \textbf{AtkMtd6} \tikz[baseline]{\node[anchor=base, draw, rounded corners] {6:14:15};}, \textbf{AtkMtd8} \tikz[baseline]{\node[anchor=base, draw, rounded corners] {1:10:16};}, \textbf{AtkMtd12} \tikz[baseline]{\node[anchor=base, draw, rounded corners] {1:10:17};} and \textbf{AtkMtd19} \tikz[baseline]{\node[anchor=base, draw, rounded corners] {3:12:16};} focus on adversarial patches, targeting camera and LiDAR systems. These attacks often involve applying adversarial stickers to road signs, causing misclassification or false detections. Attack method such as \textbf{AtkMtd5} \tikz[baseline]{\node[anchor=base, draw, rounded corners] {5:13:15};} projects fake image or signal on traffic signs to fool the camera perception. It is evident that attack medium adversarial patch is generally used in these attack methods. 

Moreover, \textbf{AtkMtd2} \tikz[baseline]{\node[anchor=base, draw, rounded corners] {2:12:15:19};}, \textbf{AtkMtd7} \tikz[baseline]{\node[anchor=base, draw, rounded corners] {7:14:15};}, \textbf{AtkMtd14} \tikz[baseline]{\node[anchor=base, draw, rounded corners] {10:16};}, and \textbf{AtkMtd20} \tikz[baseline]{\node[anchor=base, draw, rounded corners] {7:14:16/17};} involve placing physical objects (e.g., reflective boards on drones, grasses, or 3D object on roof/number plate of vehicle) to manipulate AV perception (LiDAR or radar-only perception). 3D Adversarial Object is the general attack medium for these attack methods.

Lastly, \textbf{AtkMtd9} \tikz[baseline]{\node[anchor=base, draw, rounded corners] {9:15};}, \textbf{AtkMtd10} \tikz[baseline]{\node[anchor=base, draw, rounded corners] {9:16};}, \textbf{AtkMtd11} \tikz[baseline]{\node[anchor=base, draw, rounded corners] {9:16:19};}, and \textbf{AtkMtd13} \tikz[baseline]{\node[anchor=base, draw, rounded corners] {9:17};}, \textbf{AtkMtd16} \tikz[baseline]{\node[anchor=base, draw, rounded corners] {8:15};}, \textbf{AtkMtd17} \tikz[baseline]{\node[anchor=base, draw, rounded corners] {8:15:19};}, and \textbf{AtkMtd18} \tikz[baseline]{\node[anchor=base, draw, rounded corners] {8:16};} utilize laser beams, noise, EMI and RF signals to spoof LiDAR and radar sensors, injecting false data into the perception pipeline. Similarly, \textbf{AtkMtd9} \tikz[baseline]{\node[anchor=base, draw, rounded corners] {9:15};}  exploits acoustic spoofing to mislead localization and control systems, potentially causing the vehicle to deviate from its intended path. Signal Spoofer attack medium is commonly used to implement these attack methods.

\textbf{On the Mutual Exclusivity and Completeness of Attack Methods.}
The 20 attack methods (AtkMtd1--AtkMtd20) presented in Table~\ref{tab:attack-mediums} were derived through an iterative coding process designed to balance granularity with practical utility. A single paper often describes multiple distinct attack mechanisms or targets multiple perception modules; in such cases, the paper contributes to multiple AtkMtd entries. For example, Zhu et al.~\cite{zhu2024malicious} describes attacks using reflective objects (AtkMtd14), adversarial patches (AtkMtd1), and coordinated MSF manipulation (AtkMtd20). We did not force papers into a single category; rather, each paper was coded for all applicable attack methods based on the specific mechanism described. Mutual exclusivity is maintained at the \textit{mechanism level}, not the paper level. Two attack methods are considered distinct if they differ in (i) the primary attack medium (adversarial patch vs. 3D object vs. signal spoofer), (ii) the targeted perception module, or (iii) the physical principle of exploitation. For instance, AtkMtd5 (projecting fake images onto signs) and AtkMtd1 (applying adversarial stickers) both target TSR but are distinguished by medium (signal projection vs. physical sticker). Similarly, AtkMtd10 (IEMI LiDAR spoofing) and AtkMtd18 (laser-based LiDAR spoofing) both inject phantom points but differ in the physical attack vector (electromagnetic interference vs. optical).

Completeness is necessarily bounded by the 48-paper corpus; we do not claim that these 20 methods represent all possible attack vectors. We expect new methods to emerge, particularly in underexplored areas like radar-specific adversarial perturbations and coordinated MSF attacks, and our taxonomy is designed to accommodate such extensions.

\subsubsection{Attack Goal Definitions} 
Attackers motivation to conduct various attacks can vary based on the targeted AV module. Attacks can be defined into several types based on their objectives or goals (Table~\ref{tab:Attack-def}). 
\begin{table}[hbt!]
    \centering
    \small
    \begin{adjustbox}{width=1\linewidth}
    \begin{tabular}{|>{\centering\arraybackslash}p{0.30\linewidth}|>{\raggedright\arraybackslash}p{0.60\linewidth}|} \hline
      \textbf{Attack Goal} & \textbf{Description}\tabularnewline \hline
      Hiding Attack (HA) & The attacker attempts to make a traffic sign disappear from detection\tabularnewline \hline
      Appearance Attack (AA) / Creating Attack (CA)  & The attacker introduces an object designed to be falsely recognized as a specific traffic sign\tabularnewline \hline
      Non-Target Attack (NTA) & The goal is to cause incorrect classification without targeting a specific misclassification result\tabularnewline \hline
      Target Attack (TA) & The attacker manipulates the classification so that a traffic sign is misidentified as a different, specific sign\tabularnewline \hline
      Move-In, Move-Out Attack (MIMO) & The attack shifts the object tracker’s bounding box from its original position, leading to incorrect localization\tabularnewline \hline
    \end{tabular}
    \end{adjustbox}
    \caption{Definitions and characteristics of five adversarial attack goals targeting autonomous vehicle perception systems}
    \label{tab:Attack-def}
    \vspace{-.6cm}
\end{table}

\subsubsection{Attack Classification} Different attack techniques are used to compromise AV sensors, often exploiting the fundamental operating principles of these sensors. These sensors can be exploited either from physical world defined as physical world attack or directly from inside the circuit defined as sensor attack. We have identified 6 attack classes that are classified as follows:

\textbf{Adversarial Perturbation Attack (AP):} Machine learning-based perception systems in AVs can be deceived by adversarial examples that are imperceptible to human observers but result in misclassifications or misinterpretations by the neural networks. Such adversarial manipulations can lead to critical errors such as object/lane misdetection. These perturbations facilitate attacks such as \textit{Hiding Attack (HA)}, \textit{Move-In, Move-Out Attack (MIMO)}, and \textit{Appearance Attack (AA)/Creating Attack (CA)} \cite{wang2024revisiting, zheng2024physical, cheng2022physical, chen2024adversary, sato2021dirty, jing2021too, zhao2019seeing, cheng2023fusion, zhu2021can}. \textbf{AtkMtd1}, \textbf{AtkMtd3}, \textbf{AtkMtd4}, \textbf{AtkMtd6}, \textbf{AtkMtd8}, \textbf{AtkMtd12}, and \textbf{AtkMtd19} are the methods used to conduct this class of attack.

\textbf{Physical Adversarial Object Attack (PAO):} Physical adversarial object attacks involve designing physical objects specifically to fool the AV's perception system. These objects are designed to appear benign to humans but are optimized to cause misclassification or errors in the machine learning model \cite{zhu2024ae, jia2022fooling, zhang2024online, zhu2024malicious, tu2021exploring, cao2021invisible}. \textit{HA}, \textit{AA}, \textit{TA} and \textit{NTA} attacks are conducted by them but \textit{HA} is the most prominent. \textbf{AtkMtd7} and \textbf{AtkMtd20} are found to use to conduct such attacks.

\textbf{Physical Object Attack (PO):} Physical object attacks involve placing or manipulating physical objects in the environment to disrupt or mislead the AV's perception system. These objects are not necessarily designed to be adversarial; they may simply exploit weaknesses in the system. Such as traffic cones, noise, cardboard, traffic signs, drones, rooftop cargo \cite{han2022physical, muller2022physical, lou2024first, li2023badlidet, tu2020physically}. They often create \textit{HA} and \textit{MIMO} attacks, while \textbf{AtkMtd2}, \textbf{AtkMtd14} and \textbf{AtkMtd15} are used to conduct such attacks.

\textbf{Data Poisoning Attack (DP):} During the training process, poisoned data is used to train the DNN-based object detection system. A very negligible object such as traffic cone is used to trigger the AV to take wrong perception decision \cite{han2022physical, li2023badlidet, li2023towards}. The outcome of this attack class is \textit{HA} or \textit{MIMO} and \textbf{AtkMtd2} and \textbf{AtkMtd14} are utilized to achieve the outcome.

\textbf{Denial of Service Attack (DoS):} Overloading a sensor with excessive input, such as blinding a LiDAR with high-intensity laser beams/IEMI signal or insert tracker boxes, potentially increasing the latency or processing time of the sensors \cite{ma2024slowtrack, liu2023slowlidar, jin2024phantomlidar}. This attack class causes object detection, tracking and classification failure that eventually often lead to \textit{HA}, \textit{AA}, and/or \textit{NTA} outcomes. \textbf{AtkMtd9}, \textbf{AtkMtd10}, and \textbf{AtkMtd18} are used to conduct DoS attacks.

\textbf{Signal Spoofing Attack (SS):} Attackers introduce malicious signals into the sensor's input to deceive camera, LiDAR, or radar perception \cite{sato2024invisible, cao2021invisible, jin2024phantomlidar, yang2022generating, bhupathiraju2023emi}. In this process, \textit{AA/CA}, \textit{TA}, \textit{NTA}, and \textit{MIMO} attacks are created that cause misclassification and/or misdetection of environment objects and lanes. Radar-specific spoofing attacks manipulate FMCW chirp signals or inject RF interference to create phantom obstacles or suppress real object detections \cite{sun2021control, hadad2025adversarial, komissarov2021spoofing, nallabolu2021frequency, hunt2023madradar}. Similarly, GPS spoofing manipulates positioning data, causing localization errors \cite{li2021fooling, shen2020drift}. \textbf{AtkMtd5}, \textbf{AtkMtd11}, \textbf{AtkMtd13}, \textbf{AtkMtd16}, and \textbf{AtkMtd17} are used to conduct spoofing attacks.

\begin{tcolorbox}[colback=gray!10, colframe=black, boxrule=0.4pt, arc=2pt, left=2pt, right=2pt, top=2pt, bottom=2pt]
\textbf{Insight 8: Foundational Perception Modules are Primary Attack Targets.} \textit{Early fusion attacks exploit raw multimodal inputs before they are combined, while late fusion attacks target inconsistencies after each modality is processed independently. These two failure modes expose weaknesses in both raw data coupling and decision level consistency, and they show that current fusion algorithms are not equipped to manage either type of disruption.}
\end{tcolorbox}

\begin{tcolorbox}[colback=gray!10, colframe=black, boxrule=0.4pt, arc=2pt, left=4pt, right=4pt, top=4pt, bottom=4pt]
\textbf{Insight 9: Frequently used attack methods.} \textit{Table~\ref{tab:attack-mediums} reveals that \textbf{AtkMtd18} dominates LiDAR-based attacks, appearing in eight works, and \textbf{AtkMtd2} is the most frequently used camera-oriented technique. This concentration indicates that current defenses are largely shaped by a narrow range of attack vectors, potentially overlooking a broader spectrum of underexamined threats.}
\end{tcolorbox}

\subsection{Attack Taxonomy Overview}
\label{subsec:5.3}

\begin{figure*}[hbt!]
    \centering
    \includegraphics[width=1\linewidth]{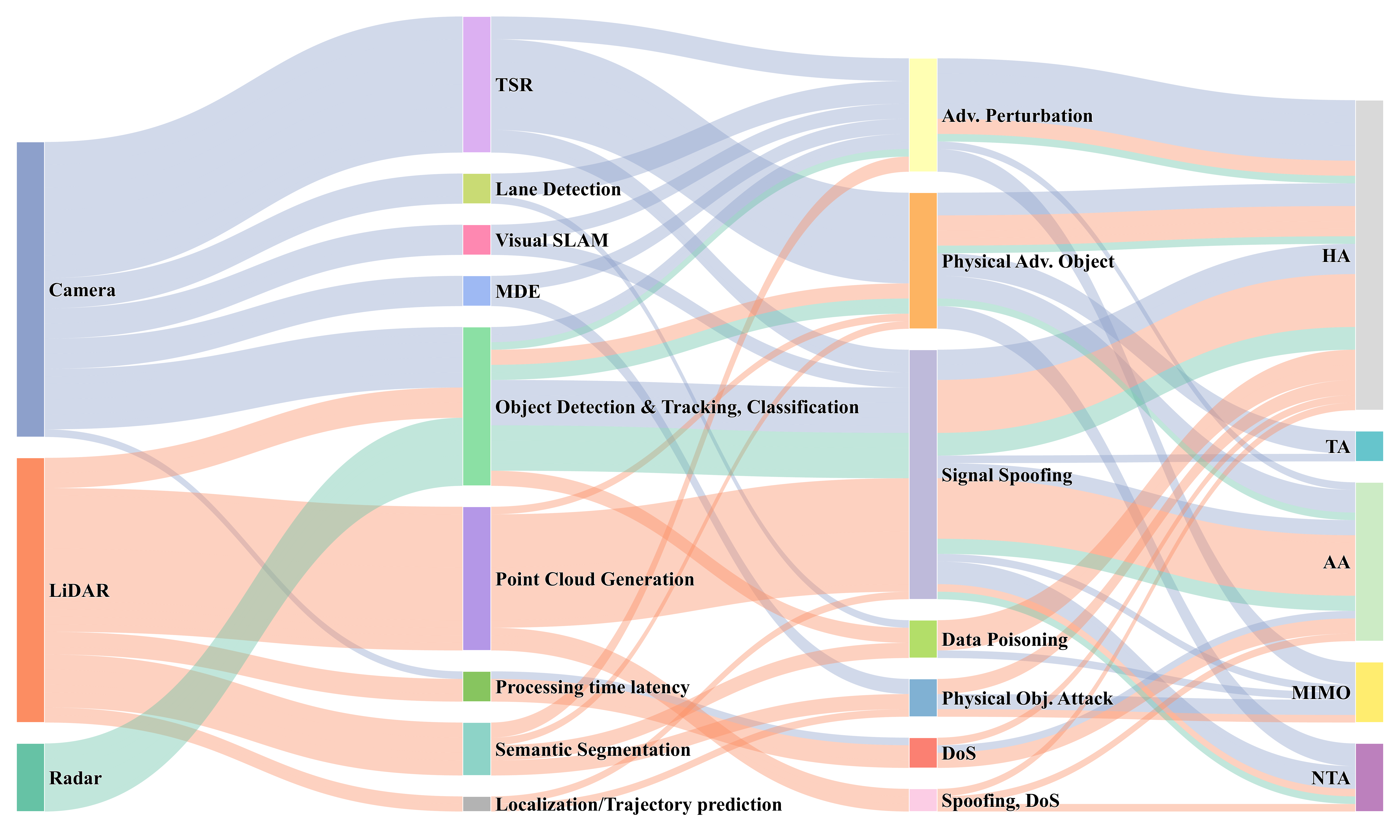}
    \caption{Taxonomic visualization of AV perception attacks. Flows represent documented causal pathways (Target AV Perception $\rightarrow$ Targeted AV Module $\rightarrow$ Attack Classification $\rightarrow$ Attack Objective). Flow thickness is proportional to the number of studies documenting each pathway. The diagram functions as a queryable map of the threat landscape.}
    \label{fig:taxonomy}
\end{figure*}

Figure~\ref{fig:taxonomy} visualizes our proposed taxonomy. Instead of presenting independent lists of attack properties, the diagram maps the actual causal paths connecting an attack vector to its final consequence. It illustrates the dependencies across four taxonomic layers: (1) \textit{Target Sensor}: the modality being attacked; (2) \textit{Target Module}: the perception subcomponent affected; (3) \textit{Attack Classification}: the mechanism of exploitation; and (4) \textit{Attacker Objective}: the intended perceptual consequence.

\paragraph{Traceability and Concentration of Risk}
Each flow in the diagram represents a documented attack path from the systematized literature, with the thickness of the line showing how much research focuses on that specific route. The main insight we gain from this visualization is the traceability and concentration of risk. It allows security engineers or researchers to look at the diagram and immediately answer practical threat-modeling questions:

\begin{itemize}
    \item \textit{``If I want to cause a Denial of Service (DoS), which sensors should I target?''} Looking at the thickest lines flowing into the DoS column shows that laser and electromagnetic interference attacks (e.g., AtkMtd9, AtkMtd10, AtkMtd16, AtkMtd18) targeting LiDAR are the most common paths.
    
    \item \textit{``If an attacker compromises the LiDAR, what is the worst they can achieve?''} Tracing the LiDAR flows left-to-right reveals that these compromises mostly lead to Object Detection and Tracking failures, which then branch into Hiding or Appearance attacks.
    
    \item \textit{``Which modules are most vulnerable to Signal Spoofing?''} The diagram shows Signal Spoofing connecting to the widest range of target modules (Object Detection, Point Cloud Generation, Classification, and vSLAM), proving it is a highly versatile attack method.
\end{itemize}

\paragraph{Key Insights and Research Gaps}
Beyond these individual paths, the visualization highlights three broader trends in the current research:

\textbf{Concentration of Research.} The thick flows starting from Camera and LiDAR show that the security community is heavily focused on visual and geometric sensors. In contrast, radar-specific attacks are still largely underexplored, even though radar is essential for driving in bad weather.

\textbf{Attack Versatility.} Signal spoofing fans out across multiple modules and objectives, making it a very flexible tool for attackers. On the other hand, physical adversarial object attacks are much more specialized, clustering tightly around object detection modules to achieve hiding or appearance goals.

\textbf{The ``Negative Space''.} Several potential attack paths are either very thin or missing entirely from the diagram. For example, there is hardly any documented work connecting radar attacks to lane detection failures, and no studies map GPS spoofing to semantic segmentation errors. These gaps do not necessarily mean the attacks are impossible; rather, they highlight blind spots in the current literature that present great opportunities for future research.
\section{Research Gaps}
\label{sec:gaps}
The systematic analysis of sensor attacks in AVs' perception reveals several critical research gaps that hinder the development of robust defense mechanisms. These gaps, identified through the examination of attack classification, attack methods, mediums, and evaluation environments, highlight the need for a paradigm shift in how the research community approaches AV security. Below, we elaborate on these gaps in detail, supported by evidence from section ~\ref{sec:current-work}.

\begin{tcolorbox}[colback=gray!10, colframe=black, boxrule=0.4pt, arc=2pt, left=4pt, right=4pt, top=4pt, bottom=4pt]
\textbf{Research Gap 1:} \textit{Despite the industry's shift toward sensor fusion, synchronous, multi-modal attack vectors represent a minimal fraction of existing research and mainly focus on manipulating a single sensor modality to disrupt MSF perception.}
\end{tcolorbox}

To address the gaps of single sensor perception systems, modern AVs rely heavily on multi-sensor fusion to enhance perception robustness. Synchronously attack vectors on MSF perception is being explored slowly, however our analysis found only 7 papers to investigate attacks on MSF systems. It is noticeable that although these attacks target the MSF perception, they targeted single modality sensor attack to disrupt the MSF perception \cite{cheng2023fusion, hallyburton2022security, shen2020drift}. In contrast, only a small subset of studies explicitly investigate attacks that simultaneously compromise multiple sensors in a coordinated/synchronized way \cite{zhu2024malicious, zhang2022play, tu2020physically, cao2021invisible}. This disconnect is alarming, as attackers could exploit inconsistencies between sensors to bypass fusion logic. For instance, an adversarial object that appears legitimate to both LiDAR and camera sensors could evade detection even if individual sensors are robust. The lack of research on cross-modal attacks leaves a critical gap in AV security.

\begin{tcolorbox}[colback=gray!10, colframe=black, boxrule=0.4pt, arc=2pt, left=4pt, right=4pt, top=4pt, bottom=4pt]
\textbf{Research Gap 2:} \textit{Existing studies predominantly focus on attacks that yield immediate perceptual effects, such as misclassification or object misdetection, while limited attention has been given to attacks that degrade perception performance progressively over time.}
\end{tcolorbox}
It is observable that the existing literature overwhelmingly prioritizes attacks with immediate consequences, such as misclassifying traffic signs \cite{wang2024revisiting} or creating phantom obstacles \cite{jin2024phantomlidar, bhupathiraju2023emi}. While these attacks demonstrate severe risks, no studies investigate gradual performance degradation. The absence of longitudinal studies on sensor performance under sustained adversarial conditions represents a significant gap, as AVs must operate reliably over years of service. Future research should model gradual degradation attacks using malware software, that would slowly corrupt LiDAR calibration parameters or degrading camera image quality or data, and develop monitoring systems to detect subtle, cumulative anomalies.
\begin{tcolorbox}[colback=gray!10, colframe=black, boxrule=0.4pt, arc=2pt, left=4pt, right=4pt, top=4pt, bottom=4pt]
\textbf{Research Gap 3:} \textit{We identified a heavy reliance on digital-world datasets such as KITTI; only a minority of studies successfully demonstrated attack feasibility under the complex, dynamic weather and lighting conditions inherent to the physical world.}
\end{tcolorbox} 
Digital-world datasets and simulations often fail to capture the full complexity of real-world driving scenarios, such as dynamic lighting, weather conditions, and unpredictable road user behavior. For example, KITTI and CARLA do not fully replicate the challenges of driving in heavy rain or fog, which can significantly impact sensor performance. Many papers rely on a small set of datasets, such as KITTI and BDD100K, which lack diversity in terms of geographic location, weather conditions, and road types. Simulation environments such as CARLA and LGSVL cannot fully replicate the physical realizability of attacks, such as the impact of adversarial patches on real-world traffic signs or the effectiveness of laser spoofing under varying atmospheric conditions. Only 15 papers include real-world testing, and even fewer test under diverse environmental conditions. This limits the generalization of attack and defense evaluations. These gaps highlight the need for more comprehensive evaluation environments that can better replicate real-world driving conditions and test the robustness of AV perception systems.

\subsection{Root Causes: Why These Gaps Persist}
\label{sec:root-causes}

The research gaps we identified are not just random omissions in the literature. They exist because of real practical and structural hurdles in the AV security field. Understanding why these gaps happen is the first step to actually fixing them.

\textbf{Cause 1: Experimental Asymmetry in Multi-Modal Attack Research.}
The lack of MSF attack research (Research Gap 1) mostly comes down to the complexity of the experiments. To attack a single sensor, an adversary only needs to bypass one input. But researchers have to perfectly synchronize attacks across multiple sensors at the exact same time to attack an MSF system. This requires expensive hardware, advanced simulators that can handle cross-sensor physics, or access to commercial AV platforms whose fusion logic is not widely accessible to academic researchers. On the other side, attacking a single sensor perception is much easier due to readily accessible tools. Due to this asymmetry, MSF attack research is comparatively underpopulated.

\textbf{Cause 2: The False Assumption that Fusion Inherently Provides Security.}
Another major reason is the underlying community assumption that multi-sensor fusion is inherently secure just because it has redundancy. We saw this often in our analysis (Insight 7): when a single-sensor vulnerability is demonstrated, the proposed mitigation is often simply to ``add fusion'' without empirical validation that the specific fusion logic can detect or withstand the demonstrated attack class. Our proof-of-concept attack (Section~\ref{sec:future}) challenges this assumption: If an attacker knows how the fusion works, they can actually use the redundancy against the system to create high-confidence fake objects. Until the community starts viewing the fusion algorithm itself as an attack surface, this gap will remain.

\textbf{Cause 3: Evaluation Infrastructure Lacking.}
The evaluation gap (Research Gap~3) persists because existing simulation platforms (CARLA, LGSVL, Apollo, AirSim) prioritize functional AV development over adversarial evaluation. These simulators lack native support of injecting physical attacks with realistic environmental settings. For example, to simulate a realistic IR laser attack, the simulator needs to understand weather interference, sensor saturation, and camera wavelengths. Simulating LiDAR spoofing requires modeling pulse timing, beam divergence, and the victim sensor's internal signal processing pipeline. Simulating coordinated MSF attacks requires synchronized injection across modalities with physically accurate sensor extrinsics. Doing this in the real world is also out of reach for most academic labs because it requires closed tracks, safety drivers, and expensive instrumented cars. Without better, security-focused simulators, it is incredibly hard to evaluate these complex attacks.

\textbf{Cause 4: Fragmented Expertise Across Disciplines.}
Studying MSF attacks requires knowledge across several different fields: signal processing (for LiDAR and radar), machine learning (for adversarial patches), and automotive engineering (to understand the fusion algorithms). Most research groups specialize in one or two of these areas, and there are few dedicated venues or workshops that explicitly bridge these disciplines in the context of AV security. This fragmentation means that a researcher capable of executing a sophisticated LiDAR spoofing attack may lack the expertise to extend it to a coordinated camera+LiDAR attack, while a machine learning researcher capable of generating adversarial patches may lack the hardware background to evaluate physical realizability.

\section{Future Directions: Exploiting Multi-Sensor Fusion Vulnerabilities}
\label{sec:future}
Our systematization reveals that while single-sensor attacks are well-studied, attacks specifically designed to exploit the fusion process in MSF systems remain relatively under-explored (Research Gap 1). The increasing reliance on MSF for robustness simultaneously introduces new vulnerabilities centered around cross-sensor inconsistencies and the fusion logic itself. Hence, we propose and validate a novel attack vector that utilizes coordinated, cross-modal manipulation to bypass typical MSF redundancy checks. 

\subsection{New Attack Vector Proposal: Combined IR Laser \& LiDAR Spoofing}

This attack aims to create a consistent, high-confidence \textit{Phantom} object detection within an early-fusion MSF system by simultaneously deceiving both the camera and LiDAR sensors in a spatially synchronized manner.

\subsubsection*{System Model}
The vehicle is equipped with a monocular camera $C$ producing image $I$ and a pulsed LiDAR $L$ producing point cloud $PC$. Each modality is processed by a modality-specific encoder $P_C$ and $P_L$ to produce low-level features $F_C$ and $F_L$. An \emph{early-fusion} module $F$ aligns these features using extrinsic calibration (e.g., $\mathbf{T}_{\text{lidar}}^{\text{cam}}$), projects or aggregates them into a common representation (commonly a Bird's-Eye-View grid), and merges them into a unified feature map that a detection head $H$ consumes to yield final 3D detections $D_{\text{fused}}$~\cite{liang2022bevfusion,yoo20203d}.

\subsubsection*{Threat Model \& Attacker Goal}
The attacker projects an IR laser pattern $\delta_I$, creating $I_{\text{adv}} = I + \delta_I$. This pattern is crafted to be interpreted by the fusion model $F$ as containing ``pedestrian-like'' visual features (e.g., texture, edges) at a specific 2D location $b_C^{\text{target}}$. Simultaneously, the attacker injects spoofed points $PC_{\text{spoof}}$, creating $PC_{\text{adv}} = PC \cup PC_{\text{spoof}}$. These points are crafted to be interpreted by $F$ as a ``pedestrian-like'' geometric feature cluster (e.g., a specific voxel pattern) at a 3D location $b_L^{\text{target}}$. The adversary's goal is to induce a high-confidence, non-existent \textit{Phantom} detection $d_{\text{fused}}\in D_{\text{fused}}$ by producing spatially consistent, sensor-level artifacts in both camera and LiDAR streams that, when aligned by $F$, form correlated features that $H$ interprets as a real object.

\subsubsection*{Attacker Capabilities}
We assume a gray-box (\LEFTcircle) attacker operating with no internal access to model weights or firmware. The attacker must have physical proximity and line-of-sight (LoS) to the target, operating at a feasible distance of $\mathcal{O}(10\!-\!100)$ meters from the victim vehicle. Their hardware capabilities include generating strong IR/optical projections to create camera-visible artifacts (using \textbf{AtkMtd5} \cite{sato2024invisible, man2020ghostimage}) and producing LiDAR spoofing returns via high-power laser or IEMI to inject spurious points into the LiDAR stream (using \textbf{AtkMtd10}/\textbf{AtkMtd18} \cite{bhupathiraju2023emi, jin2023pla, sato2023lidar}). Critically, the attacker is assumed to have coarse knowledge of the extrinsic transform $\mathbf{T}_{\text{lidar}}^{\text{cam}}$ to compute the approximate projection of a 3D spoof location into image coordinates. Finally, the attacker must be able to coordinate the camera and LiDAR artifacts with temporal alignment on the order of $\mathcal{O}(10\!-\!50)$ ms, as larger misalignments would increase the effective spatial error due to ego-motion.

\subsection{Proof-of-Concept of MSF Attack}
\subsubsection*{Simulation Assumptions}
In this paper we \emph{emulate the sensor-level outcomes} of the physical attacks rather than implement the emitting hardware: camera IR effects are modeled by perspective-aware image patch insertion and LiDAR injection is modeled by inserting synthetic point clusters at chosen 3D coordinates. 

\textbf{Formal Equivalence of Camera Simulation.}
We simulate the IR projection attack through controlled image composition where authentic object patches are inserted into target scenes. This approach is justified because both IR projections and inserted object patches create visual patterns that DNN detectors interpret as valid objects. The key security property inducing false positive detections is preserved.
Formal Equivalence: Let $I_{adv}$ represent the attacked image. In physical attack: $I_{adv} = I + \delta_{I}$, where $\delta_{I}$ is the adversarial IR projection. In our simulation: $I_{adv} = I \oplus P_{object}$, where $\oplus$ denotes composition with an object patch $P_{object}$. Both operations satisfy the critical condition: $P_C(I_{adv}) \rightarrow d_{C, \text{phantom}}$ with $confidence(d_{C, \text{phantom}}) > \tau_C$.

\subsubsection*{Physical Limitations}
This simulation abstracts away several physical factors that would affect real-world attack success and would require careful ablation in future hardware validation. Specifically, our image patch insertion does not account for:

\textit{Atmospheric effects on IR projection:} Laser power decreases over distance. At ranges of 10 to 100 meters, atmospheric scattering from humidity, dust, and ambient sunlight reduces the beam's intensity. As demonstrated in physical-world optical attacks~\cite{sato2024invisible, wang2021can}, this distance-based signal drop blurs the edges of the projected adversarial pattern, lowering the target DNN's feature extraction confidence.
    
\textit{Sensor saturation and blooming:} High-intensity laser projection can cause pixel saturation and blooming in CMOS sensors, producing artifacts that may reduce DNN confidence rather than increase it. This is discussed in prior laser attack literature, where careful power adjustment is required to avoid over-saturating the sensor~\cite{yan2022rolling, ji2021poltergeist, man2020ghostimage}.

\textit{Hardware IR Filters:} Automotive cameras are equipped with filters that block wavelengths above roughly 700~nm. To remain invisible to human drivers, an attacker must use an IR laser (e.g., 830--980~nm) that can partially bypass this filter, which requires exponentially higher wattage to register sufficient intensity on the victim sensor~\cite{sato2024invisible}. 
    
\textit{LiDAR pulse timing and receiver characteristics:} Our point cloud injection assumes perfect point insertion at arbitrary ranges; real LiDAR spoofing requires precise pulse timing to inject points into the correct range bin and must account for receiver-side processing. These challenges are well-documented in the LiDAR spoofing literature~\cite{cao2019adversarial,jin2023pla}.

Our experiments therefore validate whether \textit{plausible sensor outputs}, if produced by an attacker with sufficient hardware capability, would defeat an early-fusion perception pipeline. We do not claim end-to-end physical feasibility under all environmental conditions; rather, we demonstrate that \textit{if} an adversary can generate the modeled sensor-level artifacts, the fusion logic itself is vulnerable. Quantifying the precise conditions under which such artifacts can be physically realized, through controlled ablation of the factors listed above, represents an important direction for future hardware-in-the-loop validation.

\subsubsection*{Experimental Setup}
We extracted pedestrian image patches from KITTI validation data. These patches were inserted into target background scenes using perspective-aware transformations to maintain realistic scale/aspect ratio. Histogram matching ensured lighting/shadow consistency. The insertion location was determined by projecting the target 3D phantom location (used for LiDAR spoofing) onto the 2D image plane. We injected synthetic point clusters into the corresponding KITTI LiDAR scenes. These clusters were generated with point density and distribution statistically matching real KITTI pedestrian objects and placed at the chosen 3D phantom location. A unified script managed both the image composition and point cloud injection, ensuring the synthetic camera object and synthetic LiDAR cluster were spatially and temporally aligned according to the chosen phantom location. 
\begin{figure}[hbt!]
    \centering
    \includegraphics[width=\linewidth]{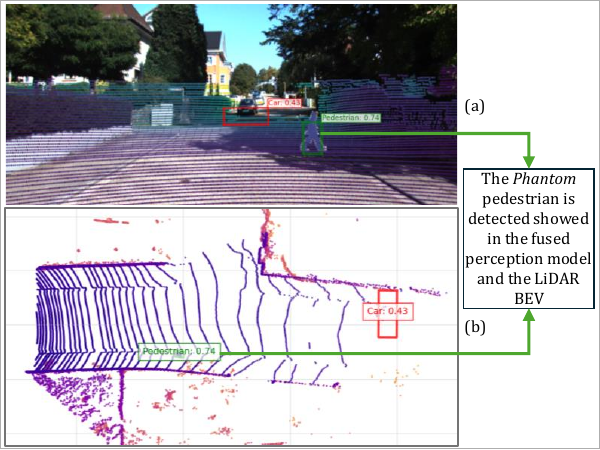}
        \caption{A representative result of a successful attack. (a) The camera-LiDAR fusion perception view, where a synthetic pedestrian patch (simulating the IR laser) and the LiDAR point cloud of the fake pedestrian (through IEMI signal) has been injected. (b) The corresponding LiDAR Bird's-Eye-View (BEV), where the spatially-aligned phantom point cloud was injected. The fusion model is successfully deceived: it correctly identifies the real `Car' (score 0.43) and, crucially, also detects the non-existent \textit{Phantom Pedestrian} (score 0.74) as a valid object.}
    \label{fig:simulation_result}
\end{figure}

\subsubsection*{Results and Analysis}
The resulting augmented data was processed by a pre-trained PointPillars~\cite{lang2019pointpillars} fusion model. The results, shown in Figure~\ref{fig:simulation_result}, demonstrate the attack's success. The model confidently identified the original car object (with 0.43 confidence score) and synthetic object as a real \texttt{Pedestrian} with a high confidence score (0.74), and the projected 3D bounding box correctly localized it within the scene. The confidence score varies with the object's distance from the camera, which is why the original car has less confidence score. This outcome validates the core premise of our proposed Attack Vector: if an adversary can successfully create cross-sensor consistency, the fusion-based perception system is highly likely to accept the phantom object as legitimate. Our simulation serves as a practical proof-of-concept, confirming that this attack vector poses a significant threat to current AV perception architectures and this attack vector is one of many other unexplored ones in the MSF perception attack spectrum.

\section{Conclusion}
\label{sec:conclusion}
This Systematization of Knowledge addressed the fragmented security landscape of AV perception through a comprehensive literature analysis. Our analysis reveals a critical research imbalance: a dominant focus on single-sensor (camera, LiDAR, radar) exploits, which masks the more systemic and underexplored vulnerabilities in MSF perception logic. Our resulting taxonomy deconstructs the threat landscape, identifying critical limitation in dynamic environment settings and the unaddressed threat of cross-modal attacks. We empirically validated the severity of these fusion-layer gaps through a proof-of-concept simulation, confirming the coordinated artifacts can successfully force high-confidence phantom detections and proving that current redundancy checks are insufficient against spatially synchronized multi-sensor threats. This SoK provides researchers and practitioners with a structured understanding of the adversarial threats facing AV perception. The identified trends, gaps, and the validated feasibility of advanced MSF attacks highlight the urgent need for the development and evaluation of robust, fusion-aware defense mechanisms capable of securing AVs against the increasingly complex and coordinated attacks of the future.

\bibliographystyle{ieeetr} 
\bibliography{references.bib}
\appendix
\section{Systematic Literature Review Details}
\label{app:systematic-review}

This appendix provides supplementary details on the systematic literature review (SLR) methodology described in Section~\ref{sec:method}. It outlines our search strategy, exclusion criteria, and the final corpus of the 48 included studies, alongside a specific justification for each paper's inclusion.

\subsection{Search Strategy and Screening Process}

As described in Section~\ref{sec:method}, we followed the five-step methodology of Wolfswinkel et al.~\cite{wolfswinkel2013using}. Initial searches across Google Scholar, IEEE Xplore, ACM Digital Library, and major security/robotics venues, combined with backward and forward snowballing, yielded 124 initial candidate papers. We utilized the following explicit search string: \textit{(``autonomous vehicle'' OR ``self-driving'') AND (``perception attack'' OR ``sensor spoofing'' OR ``adversarial attack'' OR ``LiDAR spoofing'' OR ``camera spoofing'' OR ``radar spoofing'' OR ``multi-sensor fusion attack'')}.

We screened the 124 titles and abstracts for relevance to AV perception attacks, which excluded 46 off-topic papers. The remaining 78 papers underwent a full-text review. During this phase, we applied the following inclusion criteria:
\begin{itemize}
    \item Peer-reviewed publication in a reputable conference or journal, \textit{or} a high-impact arXiv preprint (2019--2025) representing emerging work in rapidly evolving subareas.
    \item Focus on attacks targeting the perception layer of autonomous vehicles, including single-sensor and multi-sensor fusion systems.
    \item Empirical evaluation (simulation, dataset, or physical world) demonstrating attack feasibility.
\end{itemize}
During the full-text review, 30 additional papers were excluded. We explicitly excluded papers published in venues with minimal peer review (e.g., certain MDPI journals, predatory conferences), works outside the road-vehicle domain (e.g., drones, indoor robots), and purely theoretical studies without empirical validation. This highly structured process resulted in our final corpus of 48 papers.

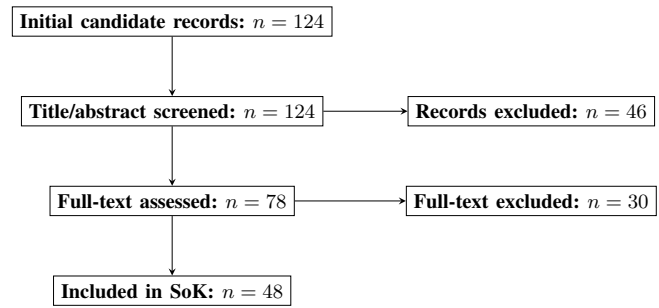
\begin{figure}[h]
\centering
\resizebox{\columnwidth}{!}{%
\begin{tikzpicture}[node distance=1.4cm, every node/.style={fill=white, font=\bfseries\small}]

\node (start) [draw, rectangle, minimum width=3.5cm] {Initial candidate records: $n=124$};
\node (screen) [draw, rectangle, below of=start, minimum width=3.5cm] {Title/abstract screened: $n=124$};
\node (ex1) [draw, rectangle, right of=screen, xshift=4.2cm, minimum width=2.5cm] {Records excluded: $n=46$};
\node (full) [draw, rectangle, below of=screen, minimum width=3.5cm] {Full-text assessed: $n=78$};
\node (ex2) [draw, rectangle, right of=full, xshift=4.2cm, minimum width=2.5cm] {Full-text excluded: $n=30$};
\node (final) [draw, rectangle, below of=full, minimum width=3.5cm] {Included in SoK: $n=48$};

\draw[->, >=stealth] (start) -- (screen);
\draw[->, >=stealth] (screen) -- (full);
\draw[->, >=stealth] (full) -- (final);
\draw[->, >=stealth] (screen) -- (ex1);
\draw[->, >=stealth] (full) -- (ex2);

\end{tikzpicture}%
}
\caption{PRISMA-style flow diagram of the paper selection process. Numbers align with Section~\ref{sec:method} methodology.}
\label{fig:prisma}
\end{figure}

Figure~\ref{fig:prisma} illustrates the complete screening flow.

\subsection{Included Papers (Final Corpus of 48 Studies)}
Table~\ref{tab:paper-list} presents the final corpus, organized by primary targeted modality (Camera, LiDAR, Radar, or Multi-Sensor Fusion). For each entry, we provide the abbreviated citation, the publication venue, and a concise justification explaining the paper's specific contribution to the SoK, whether it introduces a novel attack vector, provides foundational empirical evidence, or represents a key defensive baseline.

\begin{table*}[h]
\centering
\scriptsize
\begin{tabular}{clp{0.35\linewidth}p{0.45\linewidth}}
\toprule
\textbf{ID} & \textbf{Modality} & \textbf{Citation (Abbreviated)} & \textbf{Inclusion Justification (Primary Contribution)} \\
\midrule
\multicolumn{4}{c}{\textbf{Camera-Targeted Attacks}} \\
\midrule
1 & Camera & Wang et al., ``Revisiting Physical-World...'' NDSS '25 & Demonstrates physical patch and 3D attacks on Traffic Sign Recognition (TSR). \\
2 & Camera & Sato et al., ``Invisible Reflections,'' NDSS '25 & Introduces laser-based signal spoofing to manipulate camera TSR systems. \\
3 & Camera & Jia et al., ``Fooling the Eyes of AVs,'' NDSS '22 & Evaluates physical adversarial patches against AV object detectors. \\
4 & Camera & Chen et al., ``Adversary is on the Road,'' USENIX '24 & Exploits Visual SLAM modules using physical adversarial patches. \\
5 & Camera & Ma et al., ``SlowTrack,'' AAAI '24 & Targets processing latency (DoS) through 3D adversarial noise injection. \\
6 & Camera & Zheng et al., ``$\pi$-Jack,'' USENIX '24 & Attacks Monocular Depth Estimation (MDE) using physical adversarial objects. \\
7 & Camera & Zhao et al., ``Seeing isn't Believing,'' CCS '19 & Foundational physical adversarial patch attack on 2D object detection. \\
8 & Camera & Ji et al., ``Poltergeist,'' IEEE S\&P '21 & Explores acoustic/optical signal injection into camera hardware sensors. \\
9 & Camera & Sato et al., ``Dirty Road Can Attack,'' USENIX '21 & Uses physical dirt patterns to execute adversarial attacks on lane detection. \\
10 & Camera & Jing et al., ``Too Good to Be Safe,'' USENIX '21 & Targets lane detection systems with physical adversarial perturbations. \\
11 & Camera & Wang et al., ``I Can See the Light,'' CCS '21 & Employs laser signal spoofing to disrupt camera-based Visual SLAM. \\
12 & Camera & Man et al., ``GhostImage,'' RAID '20 & Injects phantom objects directly into camera perception pipelines. \\
13 & Camera & Muller et al., ``Physical Hijacking Attacks...'' CCS '22 & Manipulates camera object trackers using physical 3D objects. \\
14 & Camera & Han et al., ``Physical Backdoor Attacks...'' ACM MM '22 & Deploys physical backdoor triggers against camera lane detection. \\
15 & Camera & Cheng et al., ``Physical Attack on MDE,'' ECCV '22 & First physical adversarial patch attack targeting Monocular Depth Estimation. \\
16 & Camera & Zheng et al., ``Physical 3D Adv. Attacks...'' CVPR '24 & Evaluates 3D physical adversarial objects against depth estimation models. \\
17 & Camera & Yan et al., ``Rolling Colors,'' USENIX '22 & Exploits rolling shutter vulnerabilities to project adversarial color patterns. \\

\midrule
\multicolumn{4}{c}{\textbf{LiDAR-Targeted Attacks}} \\
\midrule
18 & LiDAR & Jin et al., ``PhantomLiDAR,'' NDSS '25 & Spoofs LiDAR point clouds using precise optical signal injection. \\
19 & LiDAR & Bhupathiraju et al., ``EMI-LiDAR,'' WiSec '23 & Introduces Electromagnetic Interference (IEMI) signal attacks on LiDAR. \\
20 & LiDAR & Jin et al., ``PLA-LiDAR,'' IEEE S\&P '23 & Demonstrates physical laser attacks disrupting point cloud generation. \\
21 & LiDAR & Cao et al., ``You Can't See Me,'' USENIX '23 & Explores stealthy signal spoofing to hide physical objects from LiDAR. \\
22 & LiDAR & Liu et al., ``SlowLiDAR,'' CVPR '23 & Targets LiDAR processing latency utilizing 3D adversarial objects. \\
23 & LiDAR & Zhang et al., ``An Online Defense against...'' SenSys '24 & Evaluates LiDAR object detection vulnerabilities via 3D adversarial objects. \\
24 & LiDAR & Zhu et al., ``AE-Morpher,'' USENIX '24 & Morphs 3D point clouds via physical objects to evade object detection. \\
25 & LiDAR & Sato et al., ``LiDAR Spoofing Meets...'' NDSS '24 & Analyzes next-generation LiDAR vulnerabilities to optical spoofing. \\
26 & LiDAR & Li et al., ``BadLiDet,'' TrustCom '23 & Investigates physical backdoor triggers within LiDAR object detection. \\
27 & LiDAR & Li et al., ``Towards Dynamic Backdoor...'' TrustCom '23 & Targets LiDAR semantic segmentation using dynamic 3D backdoors. \\
28 & LiDAR & Zhu et al., ``Can We Use Arbitrary Objects...'' CCS '21 & Uses everyday objects as adversarial physical perturbations against LiDAR. \\
29 & LiDAR & Cao et al., ``Adversarial Sensor Attack...'' CCS '19 & Foundational laser spoofing attack generating LiDAR phantom points. \\
30 & LiDAR & Li et al., ``Fooling LiDAR Perception...'' ICCV '21 & Spoofs LiDAR signals to actively manipulate AV trajectory prediction. \\
31 & LiDAR & Yang et al., ``Generating 3D Adv. Point Clouds,'' NDSS '22 & Generates 3D adversarial point clouds via synchronized laser injection. \\
32 & LiDAR & Zhu et al., ``Adversarial Attacks against...'' SenSys '21 & Attacks LiDAR semantic segmentation utilizing physical 3D objects. \\
33 & LiDAR & Tu et al., ``Physically Realizable Adv. Examples...'' CVPR '20 & Demonstrates physically realizable 3D adversarial objects mounted on vehicles. \\
34 & LiDAR & Lou et al., ``First Physical-World Trajectory...'' USENIX '24 & Manipulates vehicle trajectory prediction using 3D physical adversarial objects. \\

\midrule
\multicolumn{4}{c}{\textbf{Radar-Targeted Attacks}} \\
\midrule
35 & Radar & Sun et al., ``Who Is in Control?'' TIFS '21 & Spoofs FMCW radar to manipulate object detection and tracking. \\
37 & Radar & Komissarov \& Wool, ``Spoofing Attacks...'' ASHES '21 & Injects adversarial noise to spoof vehicular FMCW radar tracking. \\
37 & Radar & Nallabolu \& Li, ``A Frequency-Domain Spoofing...'' TMTT '21 & Executes frequency-domain signal spoofing on FMCW radars. \\
38 & Radar & Zhu et al., ``TileMask,'' CCS '23 & Uses physical metal-foil objects to attack radar object detection. \\
39 & Radar & Chen et al., ``MetaWave,'' NDSS '23 & Employs 3D random objects to actively disrupt radar perception pipelines. \\
40 & Radar & Hunt et al., ``MadRadar,'' NDSS '24 & Executes sophisticated signal spoofing attacks on radar velocity tracking. \\
41 & Radar & Hadad et al., ``Adversarial Attack on Automotive Radar,'' IEEE RadarConf '25 & First study on adversarial attacks targeting radar point cloud classifiers. \\

\midrule
\multicolumn{4}{c}{\textbf{Multi-Sensor Fusion (MSF) Attacks}} \\
\midrule
42 & MSF & Zhu et al., ``Malicious Attacks against MSF,'' MobiCom '24 & Demonstrates malicious cross-modal attacks manipulating fusion logic. \\
43 & MSF & Hallyburton et al., ``Security Analysis of...'' USENIX '22 & Analyzes security vulnerabilities specifically in camera-LiDAR fusion algorithms. \\
44 & MSF & Cheng et al., ``Fusion is Not Enough,'' ICLR '24 & Proves sensor fusion alone is insufficient against coordinated adversarial attacks. \\
45 & MSF & Zhang et al., ``Play the Imitation Game,'' ACSAC '22 & Executes imitation attacks targeting LiDAR-based localization modules to disrupt the fusion algorithm. \\
46 & MSF & Cao et al., ``Invisible for both Camera and LiDAR,'' IEEE S\&P '21 & Creates physical 3D objects that are stealthy to both camera and LiDAR. \\
47 & MSF & Tu et al., ``Exploring Adversarial Robustness...'' PMLR '22 & Explores the theoretical and physical robustness limits in MSF algorithms. \\
48 & MSF & Shen et al., ``Drift with Devil,'' USENIX '20 & Exploits multi-sensor fusion localization through blind-spot spoofing attacks. \\
\bottomrule
\end{tabular}
\caption{Complete list of the 48 peer-reviewed papers included in the SoK corpus, with inclusion justifications.}
\label{tab:paper-list}
\end{table*}
\end{document}